\documentclass[aps,prb,reprint,twocolumn,citeautoscript,superscriptaddress,showkeys,showpacs]{revtex4-1}
\usepackage{enumitem}
\usepackage{graphicx}
\usepackage{bm}
\usepackage{epsf}
\usepackage{ulem}
\usepackage[colorlinks = true]{hyperref}
\usepackage{amssymb}
\usepackage{amsmath}
\usepackage{graphicx}
\usepackage{rotating}
\usepackage{epsfig}
\usepackage{psfrag}
\usepackage{amsmath}
\usepackage{subfigure}
\usepackage{mathrsfs}
\DeclareMathAlphabet{\mathpzc}{OT1}{pzc}{m}{it}
\newcommand{\gf}{\mathpzc{g}}

\newcommand{\ltappr}{{{\lower2pt\hbox{$<$} } \atop \widetilde{ \ \ \ }}}
\newcommand{\gtappr}{{{\lower4pt\hbox{$>$} } \atop \widetilde{ \ \ \ }}}

\newlength{\figwidth}
\newcommand{\fg}[3]
{\begin{figure}[htb]\vspace*{-0cm}\centerline{\includegraphics[width=\figwidth]{#1}}\vskip
-0.2cm \caption{\label{#2}#3}\end{figure}}

\begin{document}
\title{Quantum Annealed Criticality:  A Scaling Description}
%:The Detailed Story
\author{Premala Chandra}
\affiliation{Center for Materials Theory, Rutgers University, Piscataway, New Jersey, 08854, USA} 
\author{Piers Coleman}
\affiliation{Center for Materials Theory, Rutgers University, Piscataway, New Jersey, 08854, USA} 
\affiliation{Department of Physics, Royal Holloway, University of London, Egham, Surrey TW20 0EX, UK}
\author{Mucio A. Continentino}
\affiliation{Centro Brasileiro de Pesquisas Fiscicas, 22290-180,
	Rio de Janeiro, RJ, Brazil}
\author{Gilbert G. Lonzarich}
\affiliation{Cavendish Laboratory, Cambridge University, Cambridge CB3 0HE, United Kingdom} 

\date{\today}

\begin{abstract}
Experimentally there exist many materials with first-order phase transitions
at finite temperature that display quantum criticality.  Classically, a
strain-energy density coupling is known to drive first-order
transitions in compressible systems, and here we generalize this Larkin-Pikin\cite{Larkin69b}
mechanism to the quantum case. We show that if the $T = 0$ system lies above
its upper critical dimension, the line of first-order transitions ends in a
“quantum annealed critical point” where zero-point fluctuations restore the
underlying criticality of the order parameter. The generalized Larkin-Pikin
phase diagram is presented and experimental consequences are discussed.
\end{abstract}
\maketitle

\section{Introduction}
The interplay of first-order phase transitions with quantum 
fluctuations 
is an active area 
\cite{Belitz99,Grigera01,Chubukov04,Belitz05,Maslov06,Rech06,Kirkpatrick12,Brando16} in the study of 
exotic quantum states near zero-temperature phase transitions
\cite{Chandra90,Chandra95b,Balents10,Zacharias15,Norman16,Paul17,Fernandes18}. 
In many metallic quantum ferromagnets, coupling of the local magnetization to 
the low energy particle-hole excitations transforms a high temperature 
continuous phase transition  into a low temperature
discontinuous one, and 
the resulting classical
tricritical points have been observed in many systems
\cite{Belitz99,Grigera01,Chubukov04,Belitz05,Maslov06,Rech06,Kirkpatrick12,Brando16}. 
Experimentally there also exist insulating materials that have 
classical first-order transitions that display quantum criticality  
\cite{Ishidate97,Suski83,Horiuchi15,Rowley14,Chandra17},  
and here we provide a theoretical basis for this behavior.
In a nutshell, we study a system with strain-energy density
coupling\cite{Larkin69b} that has a line of first-order transitions 
at finite temperatures. We show that as the temperature is lowered, 
quantum fluctuations reduce the amplitudes of
their thermal counterparts, weakening the first-order transition
and ``annealing'' the system's elastic response, 
ultimately 
resulting in a $T=0$ ``quantum annealed'' critical point. 
The generalized temperature ($T$)-tuning parameter ($g$)-field ($h$)
phase diagram
emerging from our study is presented in Figure 1 where the
field ($h$) is conjugate to the order parameter.

\figwidth=\columnwidth
\fg{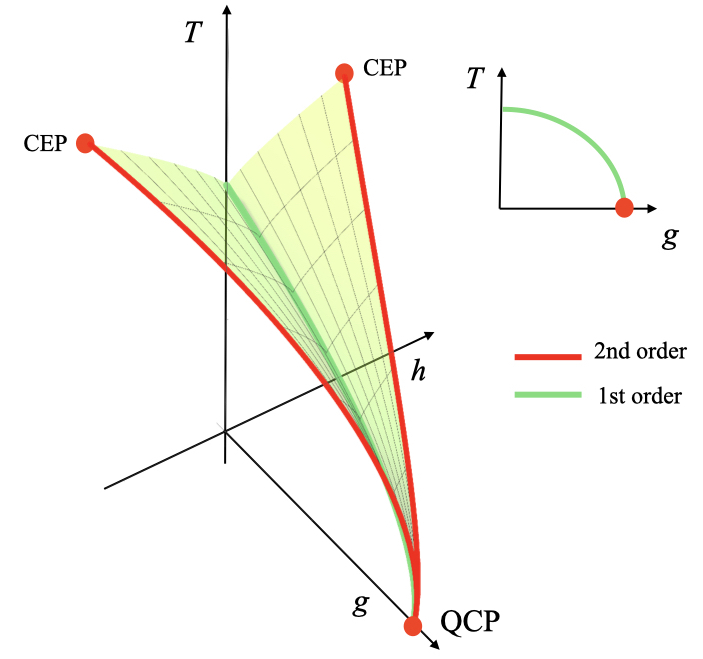}{fig1}
{Temperature ($T$)-Field ($h$)-Tuning Parameter ($g$) Phase Diagram
with a sheet of first-order transitions bounded by 
critical end-points (CEP) terminating at a zero temperature quantum critical
point (QCP); here $g$ tunes the quantum fluctuations and $h$
is the field conjugate to the order parameter.  Inset:  Temperature-Tuning Paramter ``slice'' indicating a line of classical phase transitions ending in a ``quantum annealed critical point'' where the underlying order parameter criticality is restored by zero-point fluctuations.}

At a first-order transition the quartic mode-mode coupling of 
the effective action becomes negative.
One mechanism for this phenomenon, studied by
Larkin and Pikin \cite{Larkin69b} (LP),
involves the interaction of 
strain with the fluctuating energy density of a critical order parameter. 
LP found that a diverging specific 
heat in the ``clamped'' (fixed volume)
system  leads to a first-order 
transition in the unclamped system 
at constant pressure.
The Larkin-Pikin criterion \cite{Larkin69b} for
a first order phase transition is
\begin{equation}\label{LP}
\kappa \ \ltappr  \ \frac{\Delta C_{V}}{T_{c}} \left(\frac{dT_{c}}{d \ln V} \right)^{2}
\end{equation}
where $V$ is the volume, $\Delta C_{V}$ is the singular 
part of the specific heat capacity in the clamped critical system, 
$T_{c}$ is the transition temperature
and $\frac{dT_{c}}{d {\rm ln} V}$ is its volume strain derivative. 
The effective bulk modulus $\kappa$ is defined
by $\kappa^{-1} = {\it K}^{-1} - ({\it K} + \frac{4}{3} \mu)^{-1}$
where  
${\it K}$ and $\mu$ are the bare bulk and the 
shear moduli in the absence of coupling between the order parameter
and strain;  
more physically $\kappa \sim {\it K} \frac{c_L^2}{c_T^2}$ where
$c_L$ and $c_T$ are the longitudinal and the transverse sound 
velocities \cite{Landau86}.  
An experimental setting for this behavior
is provided by $BaTiO_3$ with a classical ferroelectric phase transition
that is continuous when clamped and, due to electromechanical coupling,
becomes first-order when unclamped \cite{Borchhardt76,Lines77}. 

Low-temperature measurements on ferroelectric insulators 
provide a 
key motivation for 
our study \cite{Ishidate97,Suski83,Horiuchi15,Rowley14,Chandra17}.
At finite temperatures and ambient pressure these materials often display 
first-order transitions due to strong electromechanical 
coupling \cite{Lines77};
yet in many cases \cite{Ishidate97,Suski83,Horiuchi15,Rowley14,Chandra17} 
their dielectric susceptibilities
suggest the presence of pressure-induced quantum criticality associated with 
zero-temperature continuous
transitions \cite{Ishidate97,Suski83,Horiuchi15,Rowley14,Chandra17}. 
It is thus natural to explore whether a generalization of the
Larkin-Pikin mechanism\cite{Larkin69b} with strain-energy density
coupling,
can be developed to describe this phenomenon.

Here we generalize the Larkin-Pikin approach
to include quantum zero-point fluctuations of the energy density,
showing that it is the divergence of the energy fluctuations, both
quantum and classical, that govern the LP mechanism.
Quantum fluctuations introduce
an additional time dimension into the partition function, which now sums
over all space-time configurations\cite{sachdev_qpt_book,Millis1993}. 
At a
finite temperature $T$, the temporal extent of the quantum fluctuations
is bounded by the Planck time 
$\tau_{P}= \frac{\hbar }{k_{B}T}$ with a corresponding
quantum correlation length
$\xi_{Q}\sim (\tau_{P})^{1/z}$ 
where $z$ is the 
the dynamical exponent. Therefore for temperatures where $\xi_Q$ is greater than
the lattice spacing,
the thermal correlation volume
contains a quantum mechanical core on length- and time-scales
determined by $\xi_Q$ and $\tau_P$. 
Due to their additional time dimension, quantum fluctuations are typically
less singular than are their classical counterparts.
As the temperature is lowered, the
correlation volume of the zero-point fluctuations grows, reducing the
amplitudes of the singular thermal fluctuations 
in the clamped system.
The induced Larkin-Pikin
first order transition thus becomes progressively weaker with decreasing
temperature, leading to a continuous ``quantum annealed'' 
transition at $T=0$.

More specifically, Larkin and Pikin considered the coupling\cite{Larkin69b} 
\begin{equation}\label{lpcoupling}
{\cal L}_{I}= \lambda 
e_{ll} (\vec{x})\psi ^{2} (\vec{x}) 
\end{equation}
between the volumetric strain
field $e_{ll}$ and the squared amplitude  $\psi^{2}$ of the critical
order parameter.  In a critical system, 
the singular fluctuations of the
energy density are directly proportional to $\psi^{2}$; 
thus \eqref{lpcoupling} corresponds to a strain-energy coupling. 
Naively (\ref{lpcoupling})
is expected to induce a short-range attractive order parameter interaction.
LP showed that (\ref{lpcoupling}) also leads 
to an anomalous long-range interaction  between order parameter
fluctuations.
\begin{equation}\label{s0}
S \longrightarrow S - \frac{\lambda^{2}}{2T\kappa} 
\left[\frac{1}{V}
\int d^{3}x\int d^{3}x' \psi^{2} (\vec{x})\psi^{2} (\vec{x}') \right]
\end{equation}
with
\begin{equation}\label{}
\frac{1}{\kappa}= \left(\frac{1}{K}-
\frac{1}{K+\frac{4}{3}\mu}\right)
\end{equation}
where $\mu$ is the shear modulus.
This long-range interaction is finite if $ \mu>0 $, i.e if the medium
is a solid. 
LP showed that this induced
long-range interaction in (\ref{s0}) generates positive
feedback to the tuning parameter, leading to a multi-valued free energy
surface and a resulting first order phase transition.

Here we expand the LP approach to include both quantum and classical
fluctuations, summing over all possible spacetime configurations
in the action, to obtain a generalized LP criterion 
\begin{equation}\label{lpgeneral}
\kappa \ltappr \left(\frac{dg_{c}}{d\ln V} \right)^{2}\chi_{\psi^{2}}
\end{equation}
where
\begin{equation}\label{}
\chi_{\psi^{2}}= 
\int_{0}^{\beta }d\tau \int d^{3}x \langle \delta \psi^{2} (\vec{x})\delta \psi^{2} (0)\rangle 
\end{equation}
is the space-time average of the quantum and thermal ``energy''
fluctuations, $\beta = \frac{1}{k_{B}T}$ and
$g$ is the tuning parameter for the phase transition, with the
convention that $g_{c} (T=0)=0$.
At zero temperature, this expression extends the original LP criterion
(\ref{LP}) to
quantum phase transitions. 
At finite temperatures,  the critical
temperature and the critical coupling constant are related by 
$g_{c} (T_{c}) = uT_{c}^{1/\tilde{\Psi}}$, where $\tilde{\Psi} =\tilde{\nu} z$ is the shift
exponent governing the finite temperature transition with $\tilde{\nu}$ and z
the quantum correlation length and the dynamical critical exponents 
respectively;\cite{Continentino17} 
therefore   
$d\ln g_{c}=
\frac{1}{\tilde{\Psi} }d\ln  T_{c}$ and the LP criterion becomes 
\begin{equation}\label{}
\kappa \ltappr \left(\frac{dT_{c}}{d\ln V} \right)^{2}\overbrace
{\left( \frac{g}{2T_{c}}\right)^2\chi_{\psi^{2}}}^{\Delta C_{V}/T_{c}},
\end{equation}
where we have identified $\Delta C_{v}/T_{c}= (g/2T_{c})^{2}\chi_{\psi^{2}}$ with the specific
heat capacity.  In this way, we see that the generalized Larkin Pikin
equation encompasses the original criterion (\ref{LP}) in addition to
being
applicable at low temperatures.

Recently an adaptation of the Larkin-Pikin approach was proposed
for 
pressure($P$)-tuned quantum magnets
where it is often found
that $\frac{dT_c}{dP} \rightarrow \infty$ as $T_c \rightarrow
0$. For a pressure-tuned transition, $P-P_{c}= uT_{c}^{1/\tilde\Psi}$, so
that $dT_{c}/dP \propto T_{c}^{1-1/\tilde\Psi }$ diverges as
$T_{c}\rightarrow 0$ if $\tilde{\Psi} <1$. 
It was then argued that the associated quantum phase 
transitions be first-order \cite{Gehring08,Gehring10,Mineev17}.   
However such a diverging coupling of the 
critical order parameter fluctuations and the lattice should lead to 
structural instabilities near the quantum phase transition
that have not been observed \cite{Brando16,Bean62}.
Using Maxwell relations, we can write
$\frac{dT_c}{dP} = \left. \frac{\Delta V}{\Delta S} \right\vert_{T=T_c}$.
Since $\Delta S \rightarrow 0$ as $T_c \rightarrow 0$, proponents of the 
previous argument assume that $\Delta V$ is finite in the same limit,
indicating latent work at the quantum phase transition.  However
the ratio $\frac{\Delta V}{\Delta S}$ can still
diverge at a continous quantum transition
if the numerator and the denominator have different temperature-dependences 
as $T_c\rightarrow 0^+$.  In fact, from our generalization of the
Larkin-Pikin approach, we show that
\begin{eqnarray}\label{l}
\Delta V &\propto &- T^{\eta },\cr 
\Delta S &\propto &- T^{\eta } (T^{\frac{1}{\tilde\Psi }-1}),
\end{eqnarray}
where $\eta = \frac{{\alpha} - \tilde{\alpha}}{\alpha \tilde\Psi}$ with 
$\alpha $ and $\tilde{\alpha }$ the classical and quantum
critical exponents respectively, governing the divergence of energy
fluctuations. 
Generically ${\alpha }>\tilde{\alpha} $, since thermal fluctuations
are more singular than quantum fluctuations, so that $\eta >0$. This
means that 
\begin{equation}
\lim_{T_c \rightarrow 0^+} \Delta V \rightarrow 0
\end{equation}
so there is no latent work at the quantum phase transition, confirming
its continuous nature, despite the fact that when $\tilde{\Psi} < 1$, 
\begin{equation}
\frac{dT_c}{dP} = \left. \frac{\Delta V}{\Delta S}\right\vert_{T=T_c} \propto
-T_{c}^{1-\frac{1}{\tilde\Psi}}
\end{equation}
diverges as $T_{c}$ goes to zero. 
%For the case of interest, where $\psi = \nu z = \frac{1}{2}$, $\tilde{\alpha} =
%\frac{1}{2}$ and $\alpha = 0 $, we see that $dT_c/dP  \propto - 1/T_c$ diverges,
%$\Delta V \propto - T_c^2$
%and  

The structure of the paper is as follows.  In 
Section II we present the original Larkin-Pikin approach,\cite{Larkin69b} 
first constructing the classical LP action.
Next, following LP, we parameterize the positive feedback contribution
to the internal tuning parameter of the elastically-coupled free energy
using the uncoupled (clamped) free energy and scaling functions associated
with classical criticality.  
The non-monotonic relation between the internal tuning
parameter and the physical temperature is coincident with
a multivalued (unclamped) free energy, indicating the presence of
a first-order transition in the elastically-coupled system.  
We also derive the classical Larkin-Pikin criterion (\ref{LP}) 
as a macroscopic instability of
the original (uncoupled) critical point with respect to the strain-energy
density coupling.\cite{Bergman76}  This approach can be rewritten in terms
of correlation functions, giving insight into the $T_c \rightarrow 0$
result.  

The generalized Larkin-Pikin action is derived in Section III, where all
possible space-time configurations are summed over so that quantum and
thermal fluctuations are included.  In Section IV
a crossover scaling form for the clamped free energy that is
applicable for both the classical and quantum critical points
\cite{Continentino17} is
presented and used in the generalized Larkin-Pikin equations to study 
the system's behavior in the approach to $T_c \rightarrow 0$ along
the clamped system's critical line.  The Clausius-Clapeyron relations 
as $T_c \rightarrow 0$
are studied for the unclamped system, and it is shown that 
there is no latent work at $T_c \rightarrow 0$, confirming that the 
quantum transition is continuous. 
Next a field conjugate to the order parameter is applied in Section V,
and the critical endpoints are determined.  Field behavior in the
approach to the quantum
critical point is also studied, and these results are summarized 
in the Larkin-Pikin phase diagram.  Experimental consequences are presented
in Section VI and we end (VII) with a summarizing discussion and  open questionsfor future work. Derivations of the classical
and quantum Larkin-Pikin actions and of various crossover scaling expressions
are presented in five
Appendices for interested readers.

\section{The Classical Larkin-Pikin Approach}

The Larkin-Pikin (LP) mechanism\cite{Larkin69b} refers to a compressible
system 
where the 
order parameter, $\psi^2 (\vec{x})$, is coupled to the volumetric
strain in the simplest case of a scalar $\psi$ and isotropic
elasticity. 
The action $\mathscr {S}[\psi,u]$ for this compressible system then
divides up into three contributions
\begin{eqnarray}\label{action0}
\mathscr{S}[\psi,u] &=& \mathscr{S}_{L}[\psi] 
+ \mathscr{S}_{E}[u] + \mathscr {S}_{I}[\psi,e]  \cr
&= &\frac{1}{T}\int d^{3}x ({\cal L}_{L}[\psi]
+ {\cal L}_{E}[u]
+{\cal L}_{I}[\psi,e]).
\end{eqnarray}
Here, in order to present 
the original classical LP problem in a way that
is amenable to its quantum generalization considered later,
we have used the notation ${\cal L}$, denoting the Lagrangian
density which is also the Hamiltonian density in the classical case.

The Lagrangian density ${\cal L}_{L}[\psi ]$
describes the physics of 
the order parameter in the clamped system that, in the simplest case, is a $\psi^{4}$ field
theory
\begin{equation}\label{l3}
{\cal L}_{L}[\psi] = 
\frac{1}{2} (\partial_{\mu}\psi
)^{2}+ \frac{a}{2}\psi^{2}+ \frac{b}{4!}\psi^{4},
\end{equation}
where  $a = c (T-T_c)$ is the tuning parameter, and $ b>
0$; the clamped system thus undergoes a continuous phase transition.
The term 
\begin{equation}\label{l1}
{\cal L}_{E}[u] =
 \frac{1}{2}\left[ 
\left(K-\frac{2}{3}\mu \right)
e_{ll}^{2}+ 2\mu e_{ab}^{2}\right]- \sigma_{ab}e_{ab} 
\end{equation}
describes the elastic degrees of freedom, 
where $\sigma_{ab}$ is the external stress,
$e_{ab} (\vec{x}) = \frac{1}{2}\left(
\frac{\partial u_{a}}{\partial x_{b}}
+\frac{\partial u_{b}}{\partial x_{a}}
 \right)$ is the strain tensor, $u_{a} (\vec{x})$ is the local atomic 
displacement and 
 $e_{ll} (\vec{x})={\rm
 Tr}[e(\vec{x})]$ is the volumetric strain.  Finally
\begin{equation}\label{l2}
{\cal L}_{I}[\psi ,e]= \lambda e_{ll}\psi^{2}
\end{equation}
describes the interaction between the volumetric strain $e_{ll}$
and the squared amplitude $\psi^{2}$, the ``energy density'',
of the order parameter,
where $\lambda$ is a
coupling constant associated with the strain-dependence of $T_c$.  If
we combine 
\begin{equation}\label{}
{\cal L}_{L}+ {\cal L}_{I}= \frac{1}{2} (\partial_{\mu}\psi
)^{2}+ \frac{c}{2} (T-T_{c}[e_{ll}])\psi^{2}+ \frac{b}{4!}\psi^{4},
\end{equation}
where
\begin{equation}\label{tcstrain}
T_{c}[e_{ll}]=T_{c}- (2\lambda/c) e_{ll}
\end{equation}
is the strain dependent $T_{c}$, so that 
$(2\lambda/c) =- \left(\frac{d T_c}{d {\rm ln} V}\right)$. For
notational simplicity and convenience, we shall set
$c=1$ in the following development.

The key idea of the Larkin-Pikin approach is that we integrate out the Gaussian
strain degrees of freedom from the action 
so that the partition function
takes the form
\begin{equation}\label{summary}
Z = 
\int {\cal D}[\psi] \int {\cal D}[u ] \ e^{- \mathscr{S}[\psi,u]} 
\longrightarrow 
Z = 
\int {\cal D}[\psi]e^{- S[\psi]}.  
\end{equation}
where the effective action $S$ is a function of the order
parameter $\psi$.

Though the elastic degrees of freedom are Gaussian, and 
can be exactly integrated out,
this procedure must be done with
some care because of the special role of boundary normal modes.  In a
solid of volume  $L^{3}$, 
the normal modes can be separated into two components according to
their wavelength $\lambda$: sound waves have wavelength
$\lambda\ll L$ whereas 
boundary waves have wavelength
 comparable with
the size of the system,  $\lambda\sim L$.  
The Larkin Pikin effect is a kind of
{\sl ``elastic anomaly''}, whereby the integration over boundary
modes generates a nonlocal interaction between the order parameter in
the bulk action.  This elastic anomaly 
destroys the locality of the original theory, yet paradoxically, 
as a bulk term in
the action it is independent of the detailed boundary conditions. 

Larkin and Pikin chose periodic boundary conditions as the most
convenient way to integrate out the boundary modes.\cite{Larkin69b} 
In a system with periodic boundary conditions, the strain field separates
into a uniform ($\vec{q}=0$) ``boundary term'' and 
a finite-momentum ($\vec{q} \neq 0$) contribution determined 
by fluctuating atomic
displacements
\begin{equation}\label{splity}
e_{ab} (\vec{x}) = e_{ab} + \frac{1}{V}\sum_{\vec q  \neq  0}
\frac{i}{2}[ q_{a}u_{b} (\vec{q})+q_{b}u_{a} (\vec{q})]e^{i\vec{q} \cdot \vec{x} }
\end{equation}
where
$u_{a} (\vec{q})$ is the Fourier transform of $u_a(\vec{x})$ with 
discrete momenta $\vec{q}= \frac{2\pi}{L}
(l,m,n)$ with $\{l, m, n\}$  integers,
$\{a,b\} \in \{1,2,3\}$ and volume $V=L^{3}$.
Physically we can understand this separation in (\ref{splity}) 
by noting that
the strain only couples to the longitudinal modes; however
at $q=0$ there is no distinction between transverse and
longitudinal modes so this case must be treated separately
from the finite $q$ situation.
Formally, the solid forms a
3-torus, and the integral of the strain $e_{ab}$ around 
the torus defines the number of line defects enclosed by the torus, a
kind of flux, that is 
\begin{equation}\label{topoloop}
\oint e_{ab} (\vec{x})dx_{b} = e_{ab}\oint dx_{b } = b_{a}
\end{equation}
where $b_{b}$ is the Burger's vector of the enclosed defects. Thus on
a torus, the boundary modes of the strain have a topological
character. \color{black}

In order to integrate out the Gaussian strain degrees of freedom from
(\ref{action0}) to derive an effective action 
for the order parameter field in (\ref{summary}),
we write the effective action
\begin{equation}\label{Splus}
S[\psi]= S_{L}[\psi ]+ \Delta S[\psi ] 
\end{equation}
where 
$S_L [\psi] = \frac{1}{T} \int d^3x \ {\cal L}_{L}[\psi]$ from (\ref{l3}) 
and 
\begin{equation}\label{deltaS}
e^{-\Delta S[\psi ]}=  \int {\cal D}[e,u]e^{- (\mathscr{S}_{E}[u]
+\mathscr{S}_{I}[\psi,e])}.
\end{equation}
If we write the elastic action in a schematic, discretized form 
\begin{equation}\label{}
\mathscr{S}_{E}[u]+\mathscr{S}_{I}[\psi, e]= \frac{1}{2} \sum_{i,j}
u_{i}M_{ij} u_{j} + \lambda \sum_{j}u_{j} \psi_{j}^{2}
\end{equation}
then the effective action becomes simply 
\begin{equation}\label{deltaS2}
\Delta  S[\psi]  = \frac{1}{2} \ln {\rm det}[M] - 
\frac{\lambda^{2}}{2}
\sum_{i,j} \psi_{i}^{2}M^{-1}_{i,j}\psi_j^2
\end{equation}
where the second term is recognizable as an induced 
attractive interaction between the order parameter fields. 
Because of subtleties associated with the separation (\ref{splity}) of
the strain into uniform and finite $\vec{q}$ components, integration
of the elastic degrees of freedom in (\ref{deltaS}) leads
to an overall attractive interaction 
($\propto -\psi_{i}^{2}M^{-1}_{ij}\psi_{j}^{2}$) with both
short-range and infinite range components.

\begin{widetext}

%Integration over the strain fields 
%(\ref{splity}),
%gives rise to a correction to the action 
%of the $\psi $ fields,  $S [\psi]= S_{A}[\psi,a, b]+ \Delta S[\psi]$,
%where
%\begin{equation}\label{}
%e^{-\Delta S[\psi ]}=  \int {\cal D}[u]e^{- (S_{B}+S_{I})}.
%\end{equation}

Integrating over the elastic degrees of freedom
(\ref{splity}) in (\ref{deltaS}) 
we obtain the Larkin-Pikin action
\begin{equation}\label{aa2}
S [\psi] =  S_{L}[\psi,t,b^*] - \frac{\lambda^2}{2T}
\left(\frac{1}{K} - \frac{1}{K + \frac{4}{3}\mu}\right)
\left[ \frac{1}{V} \int d^{3}x \ \int d^{3}x' \ \psi^{2}(\vec{x}) \ \psi^{2}(\vec{x}') \right],
\end{equation} 
where 
\begin{align}\label{slocal}
S_{L} (\psi )= \frac{1}{T}\int d^{3}x\left[
\frac{1}{2} (\partial_{\mu}\psi
)^{2}+ \frac{t}{2}\psi^{2}+ \frac{b^{*}}{4!}\psi^{4}
 \right]
\end{align}
with a renormalized local interaction
\begin{equation}\label{rbb}
b^* = b - \frac{12\lambda^{2}}{K+\frac{4}{3}\mu}.
\end{equation}
where we have made the replacement $a \rightarrow t$  where
$t = (T - T_c)$ and  
$c = \frac{a}{T-T_c} = 1$.
\end{widetext}
The essence of the Larkin-Pikin effect is the appearance of a 
distance-independent interaction
between the energy densities of the order parameter field that appears
in (\ref{aa2}): it is this term that drives a
non-perturbative first order transition
at arbitrarily small $\lambda$.  
Since the Larkin-Pikin argument is valid for arbitrarily
small coupling $\lambda$, 
the perturbative $O (\lambda^{2})$ renormalization of the short-range interaction
in (\ref{rbb}) becomes negligibly small in this limit and can be
safely neglected. 
%In  all subsequent results the star on $b$ will be
%suppressed. 
The prefactor of the long range attractive interaction (\ref{aa2})
\begin{equation}\label{pref}
\frac{1}{\kappa}=\left(\frac{1}{K} - \frac{1}{K + \frac{4}{3}\mu}\right)
\end{equation}
has two competing terms. 
The first
is attractive ($\propto \frac{1}{K}$), resulting simply from integrating out 
the 
$q=0$ part of the
strain (\ref{splity}),  governed by the bulk modulus
$K$. 
The second results from the finite $q$ components of the strain
(\ref{splity}), but in a rather subtle fashion. The finite $q $ elastic
fluctuations, arising from longitudinal sound modes, are governed by
the elastic modulus $K+ \frac{4}{3}\mu$ 
and lead to the perturbative renormalization of $b$ in
(\ref{rbb}).   
However, the finite $q$ modes 
explicitly exclude a contribution from $q=0$. This ``bosonic hole'' 
in the longitudinal interactions gives rise to a residual 
long-range repulsion, 
resulting in the 
second term 
(
$\propto -\frac{1}{K + \frac{4}{3}\mu}$).
Remarkably, the overall prefactor of the long-range
interaction term ($\frac{1}{\kappa}$) is only non-zero for finite shear modulus
($\mu \neq 0$), indicating that the Larkin-Pikin effect only
occurs for solids and is absent for liquids.
We also note that for the clamped system
only the second contribution to (\ref{pref}) remains, leading
to a repulsive interaction; this is consistent with the
continuous transition of the clamped system.

The distance-independent interaction in (\ref{aa2})
can be written  in terms of the volume average of the energy density 
\begin{equation}\label{28}
\Psi^{2} \equiv \left[ \frac{1}{V} \int d^{3}x \ \psi^{2}(\vec{x}) \right]
\end{equation}
so that
\begin{equation}\label{bug}
S[\psi ]= S_{L} - \frac{\lambda^{2}V}{2T\kappa} (\Psi^{2})^{2}.
\end{equation}
Since $\Psi^2$  is an intensive variable, its fluctuations about its thermal average $ \langle \Psi^2\rangle $
\begin{equation}
\delta \Psi^2= \Psi^2 - \langle \Psi^2\rangle 
\end{equation}
become vanishingly small in the thermodynamic limit, $\langle (\delta
\Psi^2)\rangle\sim O (\frac{1}{V})$. 
Thus 
\begin{widetext}
\begin{equation}
\left(\Psi^2\right)^2 = \left( \langle \Psi^{2} \rangle + \delta
\Psi^2)\right)^2 = 2 \Psi^2 \langle \Psi^2\rangle - \langle
\Psi^2\rangle^2 + O ({1}/{V}), 
\end{equation}
so that we can reexpress (\ref{aa2}) as a set of self-consistent equations
\begin{eqnarray}\label{Action}
S [\psi] &=& 
\frac{1}{T}\int d^{3}x
\left[ {\cal L}_L(\psi,t) - \frac{\lambda^2}{\kappa} \langle \Psi^2 \rangle 
\ \psi^{2} (\vec{x}) \right]
+ \frac{\lambda^2 V }{2\kappa  }\langle \Psi^2 \rangle^{2}\cr
\cr
\langle \Psi^2 \rangle &=& \frac{\int d\psi \ \Psi^2 \ e^{-S_{L}[\psi]}}
{\int d\psi \ e^{-S_{L}[\psi]}}.
\end{eqnarray}
\end{widetext}
%where, for notational simplicity,  
%we have suppressed the B-subscript 
%of the Lagrangian density associated with the order parameter.
Equations (\ref{Action}) may be succintly formulated by introducing
an auxiliary ``strain'' variable
\begin{equation}\label{aux}
\phi = - \frac{\lambda \langle \Psi^2 \rangle}{\kappa}
\end{equation}
Then we may write
\begin{equation}\label{fenergy}
e^{-\frac{\tilde{F}(\phi)}{T}} = \int {\cal D} \psi \ e^{-S[\psi,\phi]}
\end{equation}
where
\begin{equation} \label{Sphi} 
S [\psi,\phi]  = \frac{1}{T} \int d^3x \left[{\cal L}_{A} (\psi,t) + \lambda \phi \psi^{2} + \frac{\kappa }{2} \phi^{2}  
\right] 
\end{equation}
that can be reexpressed as
\begin{equation}\label{Sphi2}
S [\psi,\phi] =  \frac{1}{T} \int d^3x \left[{\cal L}_{L} (\psi,t + 2 \lambda \phi)\right] + \frac{\kappa V}{2T} \phi^{2}
\end{equation}
where we see that the auxiliary variable $\phi$ shifts the ``mass''
(e.g. tuning parameter) of the order parameter by $a \rightarrow x= a+ 2 \lambda \phi$.
Self-consistency is then imposed as stationarity of the free energy
with respect to $\phi $, 
\begin{equation}\label{37}
\frac{\partial \tilde{F}[\phi]}{\partial \phi} = 0  
\quad \Longrightarrow \quad \left[ \lambda \langle \Psi^2 \rangle + \kappa \phi \right] V = 0.
\end{equation}

In the original Larkin Pikin derivation,\cite{Larkin69b} the action (\ref{Sphi}) was obtained by performing 
a Hubbard-Stratonovich transformation of the long-range
interaction (\ref{bug})
\begin{equation}\label{bug2}
- \frac{\lambda^{2}V}{2T} (\Psi^{2})^{2} \  \rightarrow \ \frac{\kappa V}{2T}\phi^{2}\ + \ \lambda  (\Psi^{2}) \phi,
\end{equation}
followed by a saddle-point evaluation of the integral over $\phi $.
Larkin-Pikin observed that main effect of the elasticity in the 
unclamped system is make a parameterized 
shift of the physical reduced temperature $t$ to a parameterized variable
$X$
\begin{equation}
t\rightarrow X = t + \lambda \phi.
\end{equation} 
Although
the phase transition of the unclamped system is
continuous for parameterized
parameter $X$, Larkin-Pikin\cite{Larkin69b} observed 
(see Section II.A) 
that the original (physical) tuning parameter $t[X]$  
becomes a non-monotonic function of
$X$, leading to a first-order phase transition at finite
temperatures. 

Subsequent authors pursued alternative approaches to the 
Larkin-Pikin criterion.\cite{Sak74,Bergman76,Bruno80}  
If, rather than integrating out the elasticity
variable $\phi $, 
one integrates out the order parameter fluctuations, this 
results in a reduction $\Delta \kappa $ in the bulk modulus 
that is proportional to energy density fluctuations.\cite{Bergman76}
When the specific heat diverges, the bulk modulus is negative, there
is a macroscopic 
instability and the system undergoes a first-order transition.  
In the next two sections, we summarize each of these approaches
to the classical Larkin-Pikin problem.

\subsection{Review of the Original Larkin-Pikin Argument}

The free energy of the clamped system is defined as
\begin{equation}\label{39}
e^{-\frac{{\cal F}(t)}{T}} = \int {\cal D} [\psi] \ e^{-\mathscr{S}_L[\psi,t]}
\end{equation}
where $\mathscr{S}_L$ is defined in (\ref{action0}) and we
have explicitly included its dependence on the 
tuning parameter $t$.
In writing \eqref{39}, we have glossed over issues of
renormalization.  In particular, self-energy corrections to the order
parameter propagators will shift the  critical value of $T_{c}$ in $t=T-T_{c}$
from its bare value $T_{c}$ to a renormalized transition temperature $T_{c}^{*}$.  All of
these renormalization effects can be absorbed into redefinitions of
the appropriate variables, in particular from now on  
we will redefine $t= T-T_{c}^{*}$ and for convenience we will drop the
asterix so that 
$T_{c}$ refers to the renormalized critical temperature.

From (\ref{fenergy}) and (\ref{Sphi}) we can write the free energy 
for our unclamped system as
\begin{equation}\label{40}
\tilde{\cal F} [\phi,t] = {\cal F} [X] + \frac{\kappa V}{2} \phi^2 
\end{equation}
where
\begin{equation}\label{41}
X  = t  +  2 \lambda \phi.
\end{equation}
indicates the shifting the of the tuning
parameter due to the presence of energy fluctuations.

Now
\begin{equation}\label{42}
\frac{1}{V} \frac{\partial \cal F}{\partial X} = \frac{\langle \Psi^2\rangle}{2}
\end{equation}
so that 
\begin{equation}\label{43}
\phi = - \frac{\lambda \langle \Psi^2\rangle}{\kappa} = 
-\frac{2 \lambda}{V \kappa} \left(\frac{\partial {\cal F}}{\partial X} \right)
\equiv  -\frac{2 \lambda}{V \kappa} {\cal F}'[X]
\end{equation}
where we have defined ${\cal F}'[X] \equiv \left(\frac{\partial {\cal F}}{\partial X} \right)$
for simplicity.
Therefore
\begin{equation}\label{44}
\tilde{\cal F} = {\cal F} [X] + \frac{2\lambda^2}{V\kappa} \left({\cal F}'[X]\right)^2
\end{equation}
and
\begin{equation}\label{45}
X  = t - \frac{4 \lambda^2}{V\kappa} {\cal F}'[X].
\end{equation}
Let us define
\begin{equation}\label{46}
\tilde{f} \equiv \frac{2 \lambda}{V \kappa} \tilde{\cal F} \   ,  \
\ \ f \equiv \frac{2 \lambda}{V \kappa} {\cal F} \ . 
\end{equation}
Then the two equations descibing the unclamped system are
%47
\begin{equation}\label{fx}
\tilde{f} = f [X] + \lambda \left(f'[X] \right)^2
\end{equation}
and
%48
\begin{equation}\label{ax}
t = X + 2 \lambda f'[X] 
\end{equation} 
which have to solved self-consistently, where $\tilde{f}$ and $f$
are the (renormalized) free energies of the unclamped and clamped
systems respectively.

To examine the consequences of these equations, we recall that 
that we can identify $a \propto \frac{T-T_c}{T_c} \equiv t$ with the
reduced temperature $t$.
In the clamped system, we assume a second-order transition
so we can write
\begin{equation}
f \propto -|t|^{2-\alpha}
\end{equation}
where $t$ is the reduced temperature and $\alpha > 0$ is the
exponent associated with the specific heat.
Since $a \propto t$, this leads to
\begin{equation}\label{fxs}
f \propto - |X|^{2-\alpha} \ \ {\rm and} \ \ f'[X] \propto -(2-\alpha)|X|^{1-\alpha} {\rm sgn} (X)
\end{equation}
and combining these results with (\ref{ax}) we obtain
\begin{align}\label{teq}
t & = X + 2 \lambda f'[X] \cr
  & = X - 2 \lambda (2-\alpha) |X|^{1-\alpha} {\rm sgn} (X).
\end{align}
where we see that there is a {\sl non-monotonic} relationship between
the physical temperature ($t$) and the parameterized variable ($X$), as shown in Figure
2, that leads to an inevitable first-order transition.

\figwidth=\columnwidth \fg{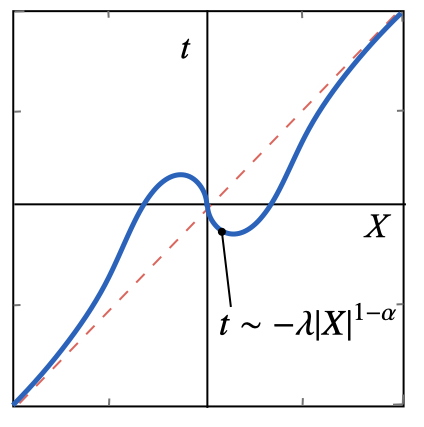}{fig1a}
{Schematic of the non-monotonic relationship between the reduced temperature ($t$) and the parameterized variable ($X$) shifted by energy fluctuations for the unclamped Larkin-Pikin problem.}

In order to see more specifically how (\ref{teq}) 
translates into a discontinuous transition let us
consider, following the example of Larkin-Pikin \cite{Larkin69b}, 
the specific case of $\alpha = 1/2$. Then for $t$ large, 
$f \propto |t|^{\frac{3}{2}}$. For $t = 0$ there are two solutions of
(\ref{teq}):  $X=0$ and $X = 4 \lambda^2$  with $f = 0$ and
$f = - \frac{16}{3} \lambda^3$ respectively.  A plot of
$\tilde{f}$ vs. $t$ is shown in Figure 3, indicating the presence
of a first-order transition.  

The Larkin-Pikin criterion (\ref{LP}) emerges from  
\begin{equation}\label{tx}
t  = X + 2 \lambda f'[X] = X + 2 \frac{\lambda^{2}}{\kappa}\langle \Psi^{2}\rangle_{X}
\end{equation} 
where again 
\begin{equation}
\langle \Psi^{2}\rangle_{X}= \frac{\int {\cal D}[\Psi ] \Psi^{2} e^{-S (X,\psi )}}
{\int {\cal D}[\Psi ]e^{-S (X,\psi )}}
\end{equation} 
is the energy density 
computed with the shifted reduced temperature $X$. This expression 
describes the relationship between the physical temperature $t =
(T-T_{c})/T_{c}$ and the parameterized variable $X$. 
Now the derivative  $dt/dX$ is given by 
\begin{equation}\label{}
\frac{dt}{dX} = 1 - \frac{\lambda^{2}V}{\kappa} \langle (\delta \Psi^{2})^{2}\rangle 
\end{equation}
We can identify the fluctuations in the right-hand side of this
equation with the specific heat capacity
\begin{equation}\label{}
C_{V}= \frac{\langle (\delta \Psi^{2})^{2}\rangle }{T_{c}^{2}}
\end{equation}
so that 
\begin{equation}\label{dtdx}
\frac{dt}{dX}= 1 - \frac{\lambda^{2}T_{c}^{2}}{\kappa }C_{V} (X)
\end{equation}
Thus if the specific heat capacity diverges at the classical critical
point of the clamped system $C_{V} (X)\rightarrow \infty $,
$dt/dX$  will change sign  as $X\rightarrow 0$. 
Then $t[X]$ becomes a non-monotonic function of
the internal temperature $X$ that  
inevitably leads to a first
order phase transition.
$\frac{dt}{dX}=0$ in (\ref{dtdx}) is 
the LP criterion (\ref{LP}).

\figwidth=\columnwidth \fg{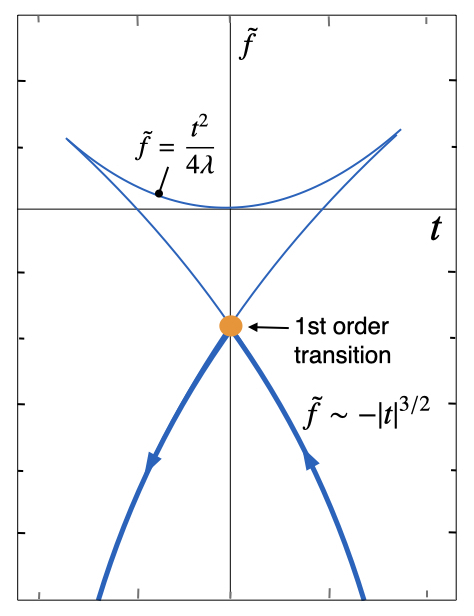}{fig1a}
{Schematic of the free energy of the unclamped compressible system
($\tilde{f}$) vs reduced 
temperature ($t$) for $\alpha = \frac{1}{2}$; the first-order transition, due 
to the non-monotonicity of $t$ vs. $X$,
is marked here.}

\subsection{The Larkin-Pikin Criterion as a Macroscopic Instability}

An alternative approach to the Larkin-Pikin criterion is 
to probe the macroscopic stability of the original critical
point with respect to the strain-energy density coupling \cite{Bergman76}.
From (\ref{fenergy}) and (\ref{Sphi})
we know that the partition function of the 
unclamped system can be written as an integral over the order
parameter fluctuations
\begin{equation}\label{}
Z[\phi] = e^{-\tilde{F}[\phi ]/T}= \int {\cal D}[\psi]\ e^{-S [\psi,\phi ]}. 
\end{equation}
The renormalized bulk modulus, 
\begin{equation}\label{rbulk}
\tilde{\kappa} \equiv \kappa - \Delta \kappa
\end{equation}
at the transition is then obtained by taking the second derivative of \eqref{fenergy} 
\begin{equation}\label{effkapclas}
\tilde{\kappa }= 
\frac{1}{V}\frac{\partial^{2}\tilde{F}}{\partial \phi^{2}} = \kappa -
\frac{\lambda^{2}} {T_c}
\int d^{3}x\langle \delta \psi^{2} (\vec{x} )\delta  \psi^{2} (0)\rangle  
\end{equation}
where 
$\delta \psi^{2} (\vec{x})   =  \psi^2(\vec{x}) - \langle \psi^2 \rangle $.
We recall the tuning parameter $a=c (T-T_{c})$ where we have set $c=1$,
so that the
singular component of the specific heat coefficient is also
proportional to the energy fluctuations
\begin{equation}\label{deltaCv}
\frac{\Delta C_{V}}{T_{c}} = \left. - \frac{\partial^{2}F}{\partial{T}^{2}} \right\vert_{{\rm Sing}}
 = \frac{1}{4T_{c}}\int d^{3}x\langle \delta \psi^{2} (\vec{x} )\delta  \psi^{2} (0)\rangle  
\end{equation}
which allows us to relate the shift in the bulk modulus to the
singular part of the specific heat
\begin{equation}\label{}
\Delta \kappa = \left( 2\lambda \right)^{2}\frac{\Delta C_{V}}{T_{c}}
 = \left(\frac{dT_{c}}{d\ln  V} \right)^{2}\frac{\Delta C_{V}}{T_{c}},
\end{equation}
where we have used \eqref{tcstrain} to identify $2\lambda =- {d T_c}/{d {\rm ln} V}$.
The condition for a macroscopic instability, and
hence a first-order transition, is
when the renormalized bulk modulus is negative
\begin{align}
\kappa - \Delta \kappa<0  \Rightarrow \ \   \kappa \ <  \  \frac{\Delta C_{V}}{T_{c}} \left(\frac{dT_{c}}{d \ln V} \right)^{2} 
\end{align}
and
we see that we have recovered the Larkin-Pikin criterion (\ref{LP}).

%\subsection{Heuristic extension to the quantum case. %Macroscopic Instability at $T=0$ ?}

The renormalization of the bulk modulus \eqref{rbulk} that results
can also be obtained diagrammatically 
(Figure \ref{fig2}).  
In this approach the bare order parameter interaction $b$ now acquires
a contribution from the coupling to the strain (Figure \ref{fig2}a).
In the Feynman diagrams $1/\kappa $ is the bare ``propagator'' for the auxilliary
strain variable $\phi $.  We can use a Dyson equation for this
strain propagator (Figure \ref{fig2}b) to determine $\tilde{\kappa}$.
Specifically we write
\begin{equation}\label{}
\left(\frac{1}{{\tilde \kappa} }\right) = \left(\frac{1}{\kappa }\right) + 
\left(\frac{1}{\kappa }\right) \lambda^2  \langle \psi^2(\vec{q}) 
\psi^2(-\vec{q}) \rangle \big\vert_{\vec{q}=0} \left(\frac{1}{{\tilde \kappa} }
\right)
\end{equation}
%that results in 
%\begin{equation}\label{LPresponse}%{\tilde \kappa} = \kappa - \Delta \kappa = \kappa - \lambda^2 
%\langle \psi^2(q) \psi^2 (-q)\rangle \vert_{q=0}.
%\end{equation}
that results in  
\begin{equation}\label{kappaeffx}
\tilde{\kappa}= \kappa - \Delta \kappa = \kappa - \lambda^{2}\chi_{\psi^{2}} 
\end{equation}
where $\chi_{\psi^{2}}$ 
is the static susceptibility for $\psi^2$, 
\begin{equation}\label{chip2}
\chi_{\psi^{2}} = \frac{1}{T_c} \int 
{d^{3}x} \langle \delta \psi^{2} (\vec{x}) \delta \psi^{2} (0)\rangle,
\end{equation}
recovering (\ref{effkapclas}). $\Delta \kappa$ is thus a self-energy
correction to the strain propagator.

\figwidth=\columnwidth \fg{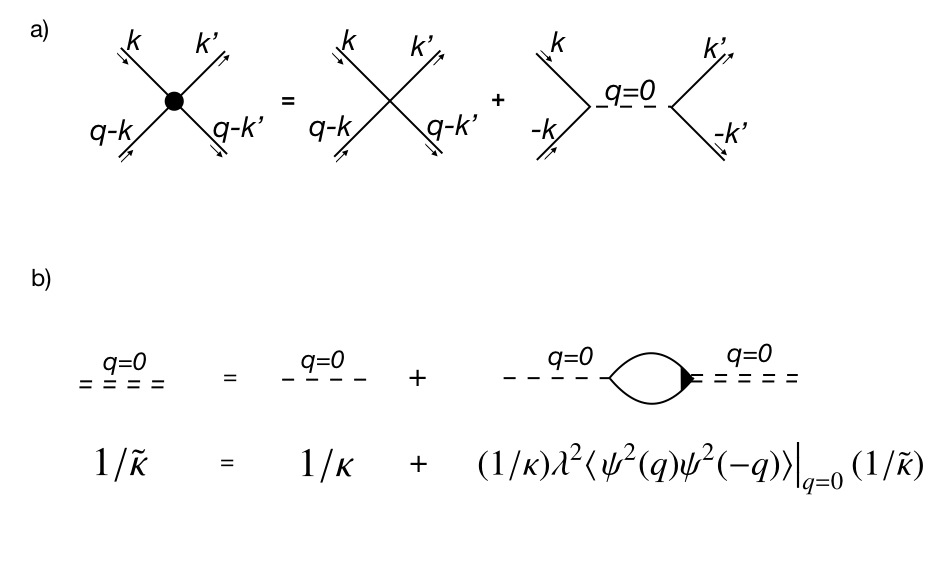}{fig2}
{Diagrammatic approach to the generalized Larkin-Pikin criterion a)
Bare interaction is a sum of a local and a nonlocal contribution
mediated by fluctuations in the strain; b) Feynman diagram showing
renormalization of the strain propagator by coupling to energy fluctuations.
}

This discussion enables us to obtain a heuristic understanding of how
the how Larkin Pikin approach can be generalized
to include quantum fluctuations of the order parameter which 
now occur in  both both space and (imaginary) time. The prefactor
$1/T_{c}$
 in
\eqref{chip2} is now replaced by an integral over time 
so that 
\begin{equation}\label{}
\chi_{\psi^{2}}
\longrightarrow
\int_{0}^{\beta } d\tau \int 
{d^{d}x} \langle \delta \psi^{2} (\vec{x},\tau ) \delta \psi^{2} (0)\rangle,
\end{equation}
where we have also generalized the expression to $d$ spatial dimensions.
This quantity is represented by the {\sl same } Feynman diagrams,
where momentum variables now become four-momenta $q= (\vec{ q},\nu_n)$. 
If we make
the Gaussian approximation $\langle \delta\psi^{2} (\vec{x})\delta \psi^{2}
(0)\rangle \approx (\langle \delta \psi (\vec{x})\delta \psi (0)\rangle
)^{2}$, then the zero-temperature limit of $\chi_{\psi^{2}}$ is
\begin{eqnarray}\label{l}
\lim_{T \rightarrow 0}\chi_{\psi^{2}} &\approx&
 \int  d\tau d^{d}x
 (\langle \delta \psi (\vec{x})\delta \psi (0)\rangle)^{2}\cr
&=&  \int \frac{d\nu}{2\pi}\frac{d^{d}q }{(2\pi)^{d}}
(\chi_{\psi} (\vec{q},\nu))^{2}
\end{eqnarray}
where in the second line we have Fourier transformed into momentum
space, and $\chi_{\psi } (\vec{q},\nu) = \langle \delta \psi (-q)\delta \psi (q)\rangle
$,  the order parameter
susceptibility, is the space-time Fourier transform of the correlator $\langle
\psi (\vec{x})\psi (0)\rangle $. 
 It follows that 
\begin{equation}
\lim_{T \rightarrow 0}
 \Delta \kappa \propto \int dq \  d\nu \ q^{d-1} \ [\chi_{\psi}(\vec{q},i\nu)]^{2}.
\end{equation}
To examine how this quantity behaves in the approach to the quantum
critical point of the clamped system, we can use dimensional
power-counting. Since $[\chi] = \left[\frac{1}{q^2}\right]$ and $[\nu]=[q^z]$,
\begin{equation*}
\lim_{T \rightarrow 0} \ [\Delta \kappa] = \frac{[q^{d+z}]}{[q^4]} \sim \xi_Q^{4- (d+z)}
\end{equation*}
where we have replaced $[q^{-1}]=[\xi_Q]$,
the quantum correlation length.  As the quantum
critical point of the clamped system is approached, $\xi_Q\rightarrow
\infty $, so that the 
quantum corrections to $\kappa$ are non-singular for $d+z>4$. 
\\

\section{Generalization of the Larkin-Pikin Approach to include Quantum  Fluctuations}

Motivated by these heuristic arguments, we now generalize the 
Larkin-Pikin approach to include both quantum and thermal
fluctuations.
In order to 
probe how the fluctuation-driven first order transition
predicted by Larkin and Pikin\cite{Larkin69b} 
evolves as the temperature is lowered to
absolute zero, we need to understand the crossover between the continuous
quantum and classical phase transitions
of the clamped system. 
From a scaling perspective, temperature is a relevant
perturbation that drives the system from a quantum to a classical
critical point. 
The action of temperature on a quantum phase transition is to introduce
a boundary condition in time, so that temperature plays the role of a 
finite-size correction at a quantum critical point.
In contrast to their static classical counterparts, quantum
zero point fluctuations are intrinsically dynamical. 
At a finite temperature $T$, the
criticality of quantum fluctuations is cut off by the Planck
time $\tau_{P}= \frac{\hbar }{k_{B}T}$, with a corresponding
quantum correlation length $\xi_{Q}\sim \tau_{P}^{1/z}$ where $z$
is the dynamical exponent; thus the static 
classical correlation volume 
contains a quantum mechanical core on length- and time-scales governed
by the Planck time.   
At low temperatures, $\xi_Q$ provides the essential short-distance cut-off 
to the static  classical fluctuations of the order parameter.  
In pictorial terms, we can visualize the fluctuations as being ``annealed'' at short distances. 

When the temperature is raised from absolute zero, there comes a point
where the finite correlation time becomes of order the Planck time,
and as the temperature is raised further, the temporal correlation
length becomes ``stuck'' at the Planck time. The temperature when this
occurs determines the quantum classical crossover.  Beyond this
point, correlations continue to grow but only in the spatial direction;
the dynamical aspect of the fluctuations is lost and
the statistical mechanics is governed purely by a sum over spatial
configurations, namely the statistical mechanics has become
classical. 

\figwidth=1.05\columnwidth \fg{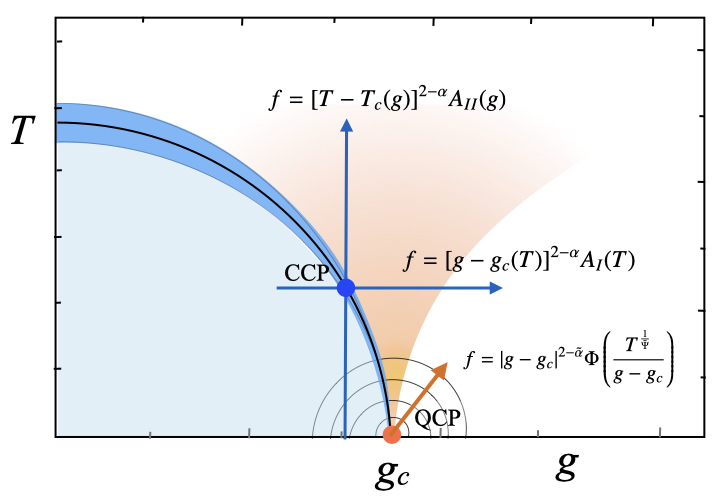}{fig4}
{Schematic showing the dependence of the Free energy of the clamped
system in the vicinity
of the quantum critical point. The scaling function about the QCP
determines the amplitude factors for the finite temperature classical
critical point (CCP), given by $A_{I} (T)$ for a constant temperature
sweep and $A_{II} (g)$ for a sweep at constant tuning parameter. Here 
the location of the quantum critical point at $g_c(0)$ is labelled as
simply $g_c$.}

More specifically near the
quantum critical point at $T=0$, the zero-point fluctuations are
governed by a finite correlation length $\xi_{Q}\sim (g-g_{c}
(0))^{-\tilde{\nu}}$, where $g$ is the parameter that tunes the transition 
and  $g=g_{c} (0)$ is the location of the quantum critical
point.  If we combine our expressions for the quantum
correlation length in the ordered phase close to the line of phase 
transitions, we find 
\begin{equation}
(g-g_c)^{-\tilde{\nu}} \sim  \left(\frac{\hbar}{k_B T_c}\right)^{\frac{1}{z}}
\end{equation}
which leads to
\begin{equation}\label{shiftex}
T_c \sim (g - g_c)^{\tilde{\nu} z} \equiv (g - g_c)^{\tilde\Psi}
\end{equation}
where $\tilde\Psi $ is called the shift exponent, and we see that 
$\Psi = \tilde{\nu} z$ if the effective
dimension of the quantum system is below its upper
critical dimension where scaling is applicable. Here we
keep with convention using this notation, hoping that there
will be no confusion with the 
the spacetime volume average of the energy density.

Larkin and Pikin\cite{Larkin69b} showed that the feedback effect of the energy
fluctuations could be reformulated in terms of the critical
temperature-dependence of the free energy of the decoupled system 
near the phase transition, allowing an analysis purely in terms of the
universal critical behavior of the decoupled system.  By generalizing 
this parameteric approach to include the effect of quantum
fluctuations, we are able to analyze the evolution of the Larkin-Pikin
system from finite to zero temperature, showing that if the energy
fluctuations are not divergent at $T=0$, the finite-temperature first-order
phase transition progressively weakens as temperature is reduced,
becoming continuous at zero temperature. 

\subsection{The Generalized Larkin Pikin Action}\label{}

The quantum mechanical action now 
picks up an additional integral over time
\begin{equation}\label{4dS}
\mathscr{S} = \int d^{4}x \ {\cal L}\equiv \int_{0}^{\beta } d\tau
\int d^{3}x \ {\cal L}.
\end{equation}
where $\beta = \frac{1}{T}$ and we recover the classical result
for large $T$.  
The spacetime generalizations of equations (\ref{l3}-\ref{l2}) 
\begin{eqnarray}\label{action0q}
\mathscr{S}[\psi,u] &=& \mathscr{S}_{L}[\psi] 
+ \mathscr{S}_{E}[u] + \mathscr{S}_{I}[\psi,e]  \cr
&= &\int  d\tau d^{3}x ({\cal L}_{A}[\psi]
+ {\cal L}_{E}[u]
+{\cal L}_{I}[\psi,e])
\end{eqnarray}
now contain kinetic energy terms, so that 
\begin{equation}\label{l3q}
{\cal L}_{L}[\psi,b] = 
\frac{1}{2} (\partial_{\mu}\psi
)^{2}+ \frac{a}{2}\psi^{2}+ \frac{b}{4!}\psi^{4}
\end{equation}
where $(\partial_{\mu}\psi)\equiv (\dot \psi )^{2}+ (\nabla\psi )^{2}
$ 
and we now identify $a = c (g-g_c^0)$, where $g_{c}^{0}$ is the bare
value of the critical coupling constant. 
The elastic degrees of freedom are now described by
\begin{equation}\label{l1q}
{\cal L}_{E}[u] =
 \frac{1}{2}\left[ 
\rho 
\dot u_{l}^{2}+ \left(K-\frac{2}{3}\mu \right)
e_{ll}^{2}+ 2\mu e_{ab}^{2}\right]- \sigma_{ab}e_{ab},
\end{equation}
and strain-energy density 
interaction 
\begin{equation}\label{l2Q}
{\cal L}_{I}[\psi,e]= \lambda e_{ll}\psi^{2}
\end{equation}
is unchanged. 
If
we combine 
\begin{equation}\label{}
{\cal L}_{L}+ {\cal L}_{I}= \frac{1}{2} (\partial_{\mu}\psi
)^{2}+ \frac{c}{2} (g-g_{c}[e_{ll}])\psi^{2}+ \frac{b}{4!}\psi^{4},
\end{equation}
where
\begin{equation}\label{tcstrain}
g_{c}[e_{ll}]=g^{0}_{c}- (2\lambda/c) e_{ll}
\end{equation}
is the strain dependent  $g_{c}$, so that 
$(2\lambda/c) =- \left(\frac{d g_c}{d {\rm ln} V}\right)$. For
notational simplicity and convenience, we shall set
$c=1$ in the following development.

Following the argument of Larkin-Pikin in the classical
case, we choose periodic boundary conditions as the most convenient
way to integrate out the elastic degrees of freedom for the analogous
quantum problem.  It is then natural to generalize the
classical expression for the strain field (\ref{splity})
to the quantum case summing over all space-time configurations
as
\begin{widetext}
\begin{equation}\label{splityQ1}
e_{ab} (\vec{x},\tau) = e_{ab}(\tau) + \frac{1}{\beta V}
\sum_{i\nu_n}
\sum_{\vec q  \neq  0}
\frac{i}{2}[ q_{a}u_{b} (q)+q_{b}u_{a} (q)]e^{i(\vec{q} \cdot \vec{x} - \nu_n\tau)}
\end{equation}
where 
$q_\alpha \equiv (\vec{q},i\nu_n)$,
$u_b(q) \equiv u_b(\vec{q},i\nu_n)$ and
$\nu_n= 2 \pi n T$ is the Matsubara bosonic frequency. 
A priori,
the  uniform strain tensor 
$e_{ab}(\tau)$ involves configurations that are time-dependent.
However, we recall that the integral of the strain field around the toroidal solid 
\begin{equation}\label{eabq1}
\oint e_{ab} (x,\tau )dx_{b} = e_{ab}(\tau) \oint dx_{b } = b_{a} (\tau)
\end{equation}
measures $b_{a}(\tau)$, the Burger's vector of the defects enclosed by the torus.  If we restrict ourselves to smooth Gaussian deformations of the solid, then the Burger's vector is a topological invariant, like the conserved winding number of superconductor.  Changes in the Burger's vector are akin to flux creep in a superconductor, and they involve the passage of dislocations across the entire solid. 
Spacetime configurations with such moving defects will be associated with large actions, making their contributions to the path integral 
exponentially small in
the thermodynamic limit.  Therefore the strain field for the
quantum Larkin-Pikin problem can be written 
\begin{equation}\label{splityQ1}
e_{ab} (\vec{x},\tau) = e_{ab} + \frac{1}{\beta V}
\sum_{i\nu_n}
\sum_{\vec q  \neq  0}
\frac{i}{2}[ q_{a}u_{b} (q)+q_{b}u_{a} (q)]e^{i(\vec{q} \cdot \vec{x} - \nu_n\tau)}.
\end{equation}

As in the classical case, our next step is to integrate out the Gaussian
elastic degrees of freedom from the action 
\begin{equation}\label{summary2}
Z = 
\int {\cal D}[\psi] \int {\cal D}[u ] \ e^{- \mathscr{S}[\psi,u]} 
\longrightarrow 
Z = 
\int {\cal D}[\psi]e^{- S[\psi]}.  
\end{equation}
where the actions now involve integrals over space-time.
We write the effective action
\begin{equation}\label{Splusq}
S[\psi]= S_{L}[\psi ]+ \Delta S[\psi ] 
\end{equation}
where 
$S_L [\psi] = \int d^4x \ {\cal L}_{L} [\psi] $ ((\ref{l3q})) 
and 
\begin{equation}\label{deltaSq}
e^{-\Delta S[\psi ]}=  \int {\cal D}[e,u]e^{- (\mathscr{S}_{e}[u]
+\mathscr{S}_{I}[\psi,e])}.
\end{equation}
Again our task is to cast this action
into matrix form
\begin{equation}\label{disq}
\mathscr{S}_{E}+ \mathscr{S}_{I}= \frac{1}{2} \sum_{q}
u_{i}M_{ij} u_{j} + \lambda \sum_{j}u_{j} \psi_{j}^{2} \ \rightarrow  \
\frac{\lambda^{2}}{2}
\sum_{i,j} \psi_{i}^{2}M^{-1}_{i,j}\psi_j^2.
\end{equation}
and to determine the nature of the induced order parameter interaction;
now the summations
run over the discrete
wavevector and Matsubara frequencies $q \equiv (i\nu_n,\vec q)$, where 
$\nu_n=\frac{2\pi }{\beta  }n$, $\vec{ q} = \frac{2\pi}{L} (j,l,k)$.
Because of the form of the strain tensor (\ref{splityQ1}), the
action in (\ref{deltaSq}) separates in two terms, corresponding
to the $q=0$ and the finite $(\vec{q},i\nu_n)$ contributions.
Integration over the elastic degrees of freedom in (\ref{splityQ1})
now results in order-parameter interactions local and nonlocal
both in space and time.

Integrating over the elastic degrees of freedom  (\ref{splityQ1}) in (\ref{deltaSq}) (see Appendix B for details), we obtain the
effective action 
\begin{equation}\label{a2q}
S [\psi] =  S^*_{L}[\psi] - \frac{\lambda^2}{2}
\left(\frac{1}{K} - \frac{1}{K + \frac{4}{3}\mu}\right)
\left[ \frac{1}{\beta V} \int d^{4}x \ \int d^{4}x' \ \psi^{2}(\vec{x}) \ \psi^{2}(\vec{x}') \right].
\end{equation}
Here 
\begin{align}\label{slocalq}
S^*_{L} (\psi )= \int d^{4}x\left[
\frac{1}{2} (\partial_{\mu}\psi
)^{2}+ \frac{\tilde{g}
}{2}\psi^{2}+ \frac{b^*}{4}\psi^{4}+ 
\frac{1}{2}\psi^{2} (\vec{x})V_{dyn} (x-x')\psi^{2} (\vec{x}') \right]
\end{align}
where $\tilde{g}= ( g-g_{c}^{0})$ and 
\begin{equation}\label{rbbq2}
b^* = b - \frac{12\lambda^{2}}{K+\frac{4}{3}\mu}
\end{equation}
is identical to that in the classical case (\ref{rbb}).
The Fourier transform of $V_{dyn}(x-x')$ is
\begin{equation}\label{vphonon}
V_{dyn} (q)= 
\frac{\lambda^{2}}{K+\frac{4}{3}\mu}\left(\frac{\nu^{2}_n/c_{L}^{2}}{\vec{q}^{2}+\nu^{2}_n/c_{L}^{2}}
\right),
\end{equation}
a dynamical order-parameter interaction where 
$q\equiv (\nu_{n},\vec{q})$ is the wavevector in
space-time. The effective action in the  quantum Larkin-Pikin problem is 
thus the sum of a
$d+z$ - dimensional generalization of the
classical effective LP action and a dynamical
interaction induced by quantum fluctuations.
Although $V_{dyn}(q)$ in (\ref{vphonon}) has $q$-dependence,
it is still local
from a scaling perspective since $V_{dyn} (q)$ is finite and non-singular 
in the limit 
$q \rightarrow 0$.  (\ref{vphonon}) 
also has the same scaling dimension as the
original local repulsive interaction in (\ref{l3q}).  
Though $V_{dyn} (q)$ does break 
Lorentz invariance, it does not reduce the order parameter symmetry.
The universality of a Wilson-Fisher fixed point is
known to be robust to such spacetime symmetry-breaking
\cite{Aharony76,Vieira2014}. For this reason the critical
behavior of the clamped system is unaffected, and thus
$V_{dyn}$ can be neglected in the local action.

We have therefore established that the generalized Larkin-Pikin action,
following the integration over the Gaussian strain including both thermal
and quantum fluctuations, is 
\begin{equation}\label{SLPqt}
S [\psi] =  S_{L}[\psi,\tilde{g},b^*] - \frac{\lambda^2}{2}
\left(\frac{1}{K} - \frac{1}{K + \frac{4}{3}\mu}\right)
\left[ \frac{1}{\beta V} \int d^{4}x \ \int d^{4}x' \ \psi^{2}(\vec{x}) \ \psi^{2}(\vec{x}') \right].
\end{equation}
with the  local action
\begin{align}\label{slocalqt}
S_{L} [\psi,\tilde{g},b^*] = \int d^{4}x \ {\cal L}_L [\psi,\tilde{g},b^*] = \int d^{4}x\left[
\frac{1}{2} (\partial_{\mu}\psi
)^{2}+ \frac{\tilde{g}}{2}\psi^{2}+ \frac{b^{*}}{4!}\psi^{4}
 \right]
\end{align}
where $b^*$ is defined in (\ref{rbbq2}).
We note that (\ref{SLPqt}) is a $d+z$-dimensional 
generalizations of the effective classical LP action, 
(\ref{aa2}), where all spacetime configurations are summed to include
both thermal and
quantum fluctuations.  Here $z$ is the dynamical
exponent associated with the temporal dimension,
since the dispersion $\omega \propto q^z$ leads to 
$[\xi_{\tau}] = [\xi]^z $ where
$\xi_\tau$ and $\xi$ are the correlation time and length respectively.

\subsection{Generalized Larkin Pikin Equations}\label{}

The development of the approach is now a simple space-time
generalization of its classical counterpart, described in equations (\ref{39}-\ref{ax}). First, 
 we perform a Hubbard-Stratonivich
transformation of the spacetime-independent interaction in (\ref{SLPqt})
\begin{equation}\label{HSqt}
- \frac{\lambda^2}{2}
\left(\frac{1}{\kappa}\right)
\left[ \frac{1}{\beta V} \int d^{4}x \ \int d^{4}x' \ \psi^{2}(\vec{x}) \ \psi^{2}(\vec{x}') \right] \ \rightarrow \ \int \ d^4x \left[(\lambda \phi) \psi^2(\vec{x})  + \frac{\kappa}{2} \phi^2\right]
\end{equation}
where
\begin{equation}\label{kappaqt}
\frac{1}{\kappa} = \frac{1}{K}- \frac{1}{K+\frac{4}{3}\mu}
\end{equation}
is the effective bulk modulus and we have introduced the auxiliary
``strain'' field $\phi$ that is spacetime independent.
Then we may write
\begin{equation}\label{partqt}
{\cal Z} = e^{-\tilde{S}(\phi)} = \int {\cal D} \psi \ e^{-S[\psi,\phi]}
\end{equation}
where $\tilde{S} = \beta \tilde{F}$ and 
\begin{equation} \label{Sphiqt} 
S [\psi,\phi]  = \int d^4x \left[{\cal L}_{L} (\psi,\tilde{g}) + \lambda \phi \psi^{2} + \frac{\kappa }{2} \phi^{2}  
\right] 
\end{equation}
that can be reexpressed as
\begin{equation}\label{Sphi2qt}
S [\psi,\phi] =  \int d^4x \left[{\cal L}_{L} (\psi,\tilde{g} + 2 \lambda \phi)\right] + \frac{\kappa V \beta }{2} \phi^{2}
\end{equation}
where we see that the auxiliary variable $\phi$ shifts the ``mass''
(e.g. tuning parameter) of the order parameter pp
by
\begin{equation} 
\tilde{g} \rightarrow X = \tilde{g} + 2 \lambda \phi.
\end{equation}
\end{widetext}

Because the second term in (\ref{Sphi2qt}) scales as the spacetime volume,
we can solve for $\phi$ using a saddle-point evaluation
\begin{equation}\label{spaqt}
\frac{\partial \tilde{F}[\phi]}{\partial \phi} = 0  
\quad \Longrightarrow \quad \left[ \lambda \langle \Psi^2 \rangle + \kappa \phi \right] V = 0
\end{equation}
where
\begin{equation}\label{edensityqt}
\Psi^{2} \equiv \left[ \frac{1}{\beta V} \int d^{4}x \ \psi^{2}(x) \right]
\end{equation}
is the spacetime volume average of the energy density
and 
\begin{equation}\label {averqt}
\langle \Psi^2 \rangle = \frac{\int d\psi \ \Psi^2 \ e^{-S_{L}[\psi]}}
{\int d\psi \ e^{-S_{L}[\psi]}}.
\end{equation}
with $S_{L}$ as in (\ref{slocalqt}).
Equations (\ref{spaqt}), (\ref{edensityqt}) amd (\ref{averqt}) 
lead to
\begin{equation}\label{94}
\phi = - \frac{\lambda}{\kappa} \langle \Psi^2 \rangle.
\end{equation}

Equations (\ref{spaqt}-\ref{94})
%(91-95) 
are identical to their classical
counterparts (\ref{28}-\ref{37}), apart from the replacement of a spatial integral by a
space-time integral in \eqref{edensityqt}.  The following development,
parameterizing the free energy of the clamped and unclamped
system, precisely follows its classical counterpart
(\ref{40}-\ref{ax}), but for completeness we include it  here in its entireity.
The free energy of the clamped
system is 
\begin{equation}\label{102}
e^{-\frac{{\cal F}(\tilde{g})}{T}} = \int {\cal D} [\psi] \ e^{-\mathscr{S}_L[\psi,\tilde{g}]}
\end{equation}
where $\mathscr{S}_L$ is defined in (\ref{action0q}) and (\ref{l3q}) and we
have explicitly included its dependence on the 
tuning parameter $\tilde{g}$.
As in the classical case, in writing \eqref{102} we have glossed over issues of
renormalization.  In particular, self-energy corrections to the order
parameter propagators will shift the quantum critical value of $g_{c}$
from its bare value $g_{c}^{0}$ to a new value $g_{c} (0)$.  All of
these renormalization effects can be absorbed into redefinitions of
the appropriate variables, in particular from now on  
we will redefine $\tilde{g}= g-g_{c} (0)$.

From (\ref{partqt}) and (\ref{Sphi2qt}) we can write the free energy 
for our unclamped system as
\begin{equation}
\tilde{\cal F} [\phi,\tilde{g}] = {\cal F} [X] + \frac{\kappa V}{2} \phi^2 
\end{equation}
where
\begin{equation}
X = \tilde{g}  +  2 \lambda \phi.
\end{equation}
indicates the shifting the of the tuning
parameter due to the presence of energy fluctuations.
Now
\begin{equation}
\frac{1}{V} \frac{\partial \cal F}{\partial X} = \frac{\langle \Psi^2\rangle}{2}
\end{equation}
so that 
\begin{equation}
\phi = - \frac{\lambda \langle \Psi^2\rangle}{\kappa} = 
-\frac{2 \lambda}{V \kappa} \left(\frac{\partial {\cal F}}{\partial X} \right)
\equiv  -\frac{2 \lambda}{V \kappa} {\cal F}'[X]
\end{equation}
where we have defined ${\cal F}'[X] \equiv \left(\frac{\partial {\cal F}}{\partial X} \right)$
for simplicity.
Therefore
\begin{equation}
\tilde{\cal F} = {\cal F} [X] + \frac{2\lambda^2}{ V\kappa} \left({\cal F}'[X]\right)^2
\end{equation}
and
\begin{equation}
X  = \tilde{g} - \frac{4 \lambda^2}{ V\kappa} {\cal F}'[X].
\end{equation}
Let us define
\begin{equation}\label{fft}
\tilde{f} \equiv \frac{2 \lambda}{V \kappa} \tilde{\cal F} \   ,  \
\ \ f \equiv \frac{2 \lambda}{ V \kappa} {\cal F} \ . 
\end{equation}
Here we recall that the integrals in the action involve
an integral over time (\ref{4dS}), $\int d^4x = \int_0^{\beta} d\tau \int d^3x $
where $\beta = \frac{1}{T}$ is a boundary term, so
that these free energies
are determined at fixed temeprature.
Therefore
the two equations describing the unclamped system are
\begin{equation}\label{fxq}
\tilde{f} = f [X,T] + \lambda \left(f'[X,T] \right)^2
\end{equation}
and
\begin{equation}\label{axq}
\tilde{g} = X + 2 \lambda f'[X,T] 
\end{equation} 
which have to be solved self-consistently.

Equation (\ref{axq}) can be rewritten as
\begin{equation}
\tilde{g} = X + \frac{2\lambda^2}{\kappa} \langle \Psi^2\rangle_X 
\end{equation}
which leads to
\begin{equation}
\frac{d \tilde{g}}{dX} = 1 - \frac{\lambda^2 V}{\kappa} 
\chi_{\psi^{2}}
\end{equation}
where 
\begin{equation}\label{}
\chi_{\psi^{2}}= 
\int_{0}^{\beta }d\tau \int d^{3}x \langle \delta \psi^{2} (\vec{x})\delta \psi^{2} (0)\rangle 
\end{equation}
is the space-time average of the quantum and thermal ``energy''
fluctuations.
Since $\frac{d \tilde{g}}{dX} = 0$ corresponds to the development
of a first-order transition, as previously discussed in the classical
case, 
analogously the generalized LP criterion is
\begin{equation}\label{lpgeneral}
\kappa \ltappr \left(\frac{dg_{c}}{d\ln V} \right)^{2}\chi_{\psi^{2}}.
\end{equation}
where we have assumed
\begin{equation}\label{gcstrain}
g_{c}[e_{ll}]=g_{c}- 2\lambda e_{ll}
\end{equation}
similar to (\ref{tcstrain}).
At zero temperature, this expression generalizes the original LP criterion
(\ref{LP}) to
quantum phase transitions. 
At finite temperatures,  the critical
temperature and the critical coupling constant are related by 
$g_{c} (T_{c}) = uT_{c}^{1/\tilde{\Psi}}$, so that    
$d\ln g_{c}=
\frac{1}{\tilde{\Psi} }d\ln  T_{c}$ and the LP criterion becomes 
\begin{equation}\label{}
\kappa \ltappr \left(\frac{dT_{c}}{d\ln V} \right)^{2}\overbrace
{\left( \frac{g}{2T_{c}}\right)^2\chi_{\psi^{2}}}^{\Delta C_{V}/T_{c}},
\end{equation}
where we have identified $\Delta C_{v}/T_{c}= (g/2T_{c})^{2}\chi_{\psi^{2}}$ 
with the specific
heat capacity.  Thus we see that the generalized Larkin Pikin
equation encompasses the original LP criterion, (\ref{LP}) and also
(\ref{dtdx}),in addition to
being applicable at low temperatures.
Our next step is to identify a crossover scaling
form for the clamped free energy, $f$, that 
includes both thermal and quantum critical fluctuations.

\begin{widetext}

\section{Quantum Annealing of the First-Order 
Transition}

\subsection{The Amplitude Factors}

In order to generalize the Larkin-Pikin argument to $T \rightarrow 0$,
we need to introduce a crossover scaling form for the clamped
free energy $f$ in (\ref{fxq}) and (\ref{axq}) that is applicable near both
the classical and the quantum critical points.
The approach we follow here that describes both the quantum and 
classical cases \cite{Continentino17} 
was adapted from an earlier study
used to describe Ising anisotropy at a Heisenberg 
critical point.\cite{pfeutyfisher74}

At a finite temperature, the location of the phase transition  is
shifted by the thermal fluctuations, so that 
\begin{equation}\label{a1}
g_{c} (T) = g_{c} (0) - uT^{\frac{1}{\tilde\Psi }}
\end{equation}
where $\tilde\Psi$ is the shift exponent defined in (\ref{shiftex}); we note that if the effective
dimension of the quantum system is at or below its upper
critical dimension $\tilde\Psi = \tilde\nu z$. 
For convenience, we will shift the definition of $g$ to absorb 
the zero temperature QCP critical coupling
constant,  $g_{c} (0)$, i.e $g-g_{c} (0)\rightarrow g$, so that $g_{c}
(T)= - uT^{\frac{1}{\tilde{\Psi }}}$. 
Now temperature is a finite
size correction to the quantum critical point,  and the 
free energy is determined by a crossover function 
\begin{equation}\label{a2}
f (g,T)= g^{2-\tilde{\alpha}} \Phi \left(\frac{T^{\frac{1}{\tilde\Psi }}}{g}\right).
\end{equation}
which describes both the quantum critical
point, and the finite temperature classical critical point of the
clamped system (see Figure 5), here we will use the convention
that an exponent with a tilde refers to the quantum case so that
$\alpha$ and $\tilde{\alpha}$ are classical and quantum exponents 
respectively.  
%Mucio's sentence goes here
A key point is that at finite temperature, critical
behavior now occurs at the shifted value of $g_{c} (T)$, and the
scaling behavior is governed by the finite temperature critical exponents. 
Therefore for a fixed temperature 
scan (Fig. \ref{fig4}) for small $g-g_{c} (T)$, 
\begin{equation}\label{a3}
f (g,T)= (g-g_{c} (T))^{2-{\alpha }} A_I (T).
\end{equation}
where $A_I(T)$ is the amplitude factor for the classical critical
point occuring at $g=g_{c} (T)$.
Similarly if we perform a sweep through the phase transition at
constant coupling constant $g$ (Fig. \ref{fig4}), 
then we can write
\begin{equation}\label{xover2a}
f[g,T]\sim (T-T_{c}[g])^{2-{\alpha }}A_{II} (g),
\end{equation}
where $A_{II}(g) $ is amplitude factor for the quantum transition 
at $T_{c}[g]= (-g/u)^{\tilde{\Psi }} $. 
The scaling form (\ref{a2}) allows us to determine the form of these
amplitude factors (see Appendix C), given by 
\begin{eqnarray}\label{ampfactors}
A_{I} (T)&=& a_{1}T^{\left(\frac{{\alpha }-\tilde\alpha }{\tilde\Psi
} \right)},\cr\cr
A_{II} (g)&=& a_{2}g^{(1-\tilde\Psi ) (2-{\alpha })+ ({\alpha}-\tilde\alpha )},
\end{eqnarray}
where $a_{1}$ and $a_{2}$ are constants. 
The resulting expressions for the singular parts of the
free energy for constant temperature and constant coupling constant 
sweeps (see Figure \ref{fig4}) are  
\begin{eqnarray}\label{summaryX}
f[g,T]=\left\{\begin{array}{rlr}
 |g-g_{c} (T) |^{2-{\alpha }}&
T^{
\frac{{\alpha}-\tilde{\alpha}}{\tilde \Psi }
}
&
\hbox{(constant T)},
\cr\cr
 |T-T_{c}[g]|^{2-{\alpha }}&
g^{(1-\tilde\Psi ) (2-{\alpha })+ ({\alpha}-\tilde\alpha )}&
\hbox{(constant g)}
,
\end{array}
\right.
\end{eqnarray}
where, since we are interested in the singular scaling behavior, we
have dropped the constants $a_{1}$ and $a_{2}$. 
\end{widetext}

\subsection{Clausius-Clapeyron Relations as \texorpdfstring{$T_c \rightarrow 0$}{Tc->0}}

We now examine how the discontinuities $\Delta S (T_{c})$ and $\Delta
V (T_{c})$ in entropy and volume  evolve along the first order phase
boundary as the transition $T_{c}$ is lowered towards zero, and connect them with 
the Clausius-Clapeyron relation.  
In this discussion, we shall identify the tuning
parameter ${g}$ with the pressure $P$, ${g}\equiv P-P_{c}$.
Using Maxwell's relations we have
\begin{equation}\label{CC0}
\frac{dT_c}{dP_{c}} \equiv 
\frac{dT_c}{d{g_{c}}}=
 - \left. \frac{\Delta V}{\Delta S} \right\vert_{T=T_c}
%=  \frac  {\frac {d\tilde{f}}{d{g}}   }  {\frac{d\tilde{f}}{dT_{c}}}.
\end{equation}
From (\ref{a1}), we have
\begin{equation}\label{CCdirect}
\frac{dT_c}{d{g_{c}}} \propto - T_c^{1 - \frac{1}{\Psi}}.
\end{equation}
In the case of particular interest, that of three-dimensional
ferroelectrics, the dynamical exponent $z=1$, so the effective
dimension $d_{eff}= 3+z=4$ lies at the upper critical dimension. 
In this case, $\tilde\Psi = \tilde{\nu} z = \frac{1}{2}$, we see that
this $dT_{c}/dP_{c}\propto T_{c}^{{-1}}$, 
implying that $\Delta V/\Delta S$ diverges as $T_{c}\rightarrow 0$. 
To understand how this happens, 
we now independently evaluate 
the temperature-dependences of $\Delta V$ and $\Delta S$.

To carry out this calculation, we need to 
input the quantum-renormalized amplitude factors for the free energy 
into the parameterized equations (\ref{fxq})  and (\ref{axq}). We
consider tuning through the first order phase transition at constant $T=T_{c}$.
The corresponding tuning variable $\gf=g-g_{c} (T_{c})$ of the clamped system 
in \eqref{summaryX}  is now replaced by
the parametric variable $X$ describing the tuning parameter that has
been shifted by the long range interactions, $\gf\rightarrow X $. 
The singular part of the free energy, by (\ref{summaryX}),
is then
\begin{equation}\label{qlpz1}
f[X]= - |X|^{2-{\alpha} }T_{c}^{
\frac{{\alpha}-\tilde{\alpha}}{\tilde\Psi  }}. 
\end{equation}
Using (\ref{fxq})  and (\ref{axq}),
the explicit form of the quantum Larkin-Pikin equations are then 
\begin{equation}\label{qlp1}
\tilde{f}[X] = -|X|^{2-{\alpha}} 
T_{c}^{
\frac{{\alpha}-\tilde{\alpha}}{\tilde\Psi  }}
+ \lambda \left [ (2-{\alpha}) 
X^{1-{\alpha}}
T_{c}^{
\frac{{\alpha}-\tilde{\alpha}}{\tilde\Psi }}
\right]^2
\end{equation}
and 
\begin{equation}\label{qlp2}
{\gf}[X] =  X - 2\lambda (2 - {\alpha}) |X|^{1-{\alpha}}
T_{c}^{
\frac{{\alpha}-\tilde{\alpha}}{\tilde\Psi  }}
 {\rm sgn}(X).
\end{equation}
The only difference between this calculation and the original Larkin
Pikin calculation is the presence of the amplitude factors $T_{c}^{
\frac{
{\alpha }-\tilde{\alpha} 
}{\tilde\Psi }
}$.
From the original Larkin Pikin analysis, we know that since
${\gf}[X]$ is a non-monotonic function of $X$,  the inverse function
$X[\gf]$ is a discontinuous function of $\gf$, given by $X[\gf] = {\rm sgn } (\gf) X_{+} (|\gf|)$.
In particular, at ${ \gf}=0$, $X$ jumps from $-X_{c}$ to $+X_{c}$,
Since the free energy $\tilde{f}[X_{c}]=\tilde{f}[-X_{c}]$ is an even
function of $X$,  it follows that the first order transition occurs at
${\gf}=0$. 
Using 
${\gf}[X_{c}]=0$,  we obtain
\begin{equation}\label{xgc}
X_{c} = [2\lambda \ (2 - {\alpha})]^{\frac{1}{{\ }}}
T_c^{\frac{{\alpha}-\tilde{\alpha}}{\alpha \tilde\Psi }}.
\end{equation}

To obtain 
$\Delta V= d\tilde{f}/d{\gf}= \tilde{f}'[X_{c}]/{\gf}'[X_{c}]
 $, we need ${\gf}'[X]$ and 
$\tilde{f}'[X]$.  First, we differentiate \eqref{qlp2} with respect to
$X$, and using the expression \eqref{xgc} for $X_{c}$, we find
${\gf}'[X_{c}] = {\alpha } $ is just a constant. 
Also, differentiating \eqref{qlp1} with respect to $X$ and substituting
\eqref{xgc}, we find that 
\begin{equation}
\tilde{f}'[X] = - ({\alpha }/2\lambda)X_{c}.
\end{equation}
from which we obtain 
\begin{equation}\label{deltav}
\Delta V (T_{c}) \propto - T_c^{\frac{{\alpha} - \tilde{\alpha}}{{\alpha}\tilde\Psi}}.
\end{equation}
Similarly, to obtain $\Delta S = - d\tilde{f}/dT_{c} = -\tilde{f}'[X]
dX/dT_{c}$,we need $dX/d{T_{c}}$.  Now since ${\gf}=g+uT_{c}^{1/\tilde\Psi}$ and
${\gf}'[X_{c}]={\alpha }$, 
we obtain
\begin{equation}
\frac{dX}{dT_c} = \frac{1}{\gf'[X_{c}]}\frac{d{\gf}}{dT_{c}}=
\frac{u}{
 {\alpha }\tilde\Psi}T_c^{\frac{1}{\tilde\Psi} - 1}.
\end{equation} 
so that 
\begin{equation}\label{deltat}
{\Delta S}[T_{c}] = -\tilde{f}'[X_{c}]\frac{dX}{dT_{c}}
\propto T_c^{\frac{{\alpha} - \tilde{\alpha}}{{\alpha}\tilde\Psi}}
T_c^{\frac{1}{\tilde\Psi} - 1}.
\end{equation}

For the case $\tilde\Psi = \tilde\nu z=1/2$, ${\alpha }=1/2$, $\tilde{\alpha}=0$, 
both $\Delta V\sim T_{c}^{2}$ and $\Delta S\sim T_{c}^{3}$
vanish at
absolute zero, but in such a way that their ratio diverges
as $T_{c}\rightarrow 0$, in agreement with 
(\ref{CCdirect}). Naively, the divergence 
of $\Delta V/\Delta S$ as  $T_{c}\rightarrow 0$ might be taken as
evidence that the tendency towards a first order transition increases
as the temperature goes to zero, yet the paradox is resolved by noting
that $\Delta S$ and $\Delta V$ 
simply vanish at different rates, still signifying
an approach to a continuous quantum phase transition.  More generally, so
long as the finite temperature 
exponent ${\alpha }$ exceeds the quantum
exponent $\tilde{\alpha} $, ${\alpha }>\tilde{\alpha} $, 
(\ref{deltav}) indicates that 
\begin{equation}
\lim_{T_c \rightarrow 0} \Delta V \rightarrow 0
\end{equation}
so that quite generally, there is no latent work as $T_c$ goes to zero, indicating
that quantum fluctuations ``anneal''
the zero-temperature quantum phase transition to become continuous
(see Figure \ref{Fig6}).

\onecolumngrid
\begin{center}
\figwidth=\textwidth\fg{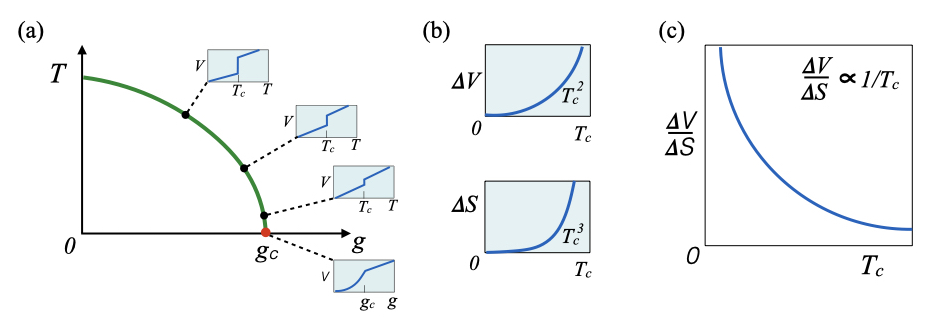}{Fig6}{Schematic figure
showing the evolution of the first order phase transition in the
approach to the quantum annealed critical point for the case
$\tilde\Psi = \tilde\nu z=1/2$, ${\alpha }=1/2$, $\tilde{\alpha}=0$. (a) Evolution of
jump in volume (b) dependence of $\Delta V$ and $\Delta S$ on $T_{c}$
and (c) $T_{c}$ dependence of $\Delta V/\Delta S$.}
\end{center}
\twocolumngrid

\section{The Larkin-Pikin Phase Diagram}

We therefore have a system with a line of classical
first-order transitions that ends in a $T=0$ quantum critical point.
Next we consider application of a field conjugate and parallel/antiparallel
to the order parameter.  In this Section we  present the 
scaling approaches to the critical endpoints, classical and quantum, 
and the resulting temperature-field-quantum
tuning parameter ($g$) phase diagram of the generalized Larkin-Pikin problem.

\subsection{Identification of the Classical Critical Endpoints}

We can work out the scaling of the critical end point in the LP
mechanism using the scaling form for the free energy
\begin{equation}\label{fhscaling}
f\propto - t^{2-\alpha } \Phi\left(\frac{h}{t^{\beta \delta }} \right).
\end{equation}
where
$h$ is the dimensionless external field,  
$t= (T-T^0_{c})/T^0_{c}$ is the reduced temperature and
$T_c^0$ is the transition temperature of the clamped system.
We want to know how (\ref{fhscaling}) behaves in a finite field when
$t$ is small compared with $h^{1/\beta \delta }$.  
In this limit we know that $f\propto h^{\frac{1}{\delta }+1}$ and 
(\ref{fhscaling}) 
can be rewritten
\begin{equation}\label{fhscaling2}
f\propto - h^{\frac{1}{\delta }+1}\Lambda\left(\frac{t}{h^{1/\beta
\delta }} \right)= 
-
 h^{\frac{2-\alpha}{\beta \delta }} 
\Lambda\left(\frac{t}{h^{1/\beta\delta }} \right)
\end{equation}
where,using the identity $2-\alpha = \beta (1+\delta )$, 
we have substituted $(\delta +1)/\delta  = (2-\alpha )/\delta \beta $,
Comparing (\ref{fhscaling}) and (\ref{fhscaling2}), we see
that 
\begin{equation}\label{lambdadef}
\Lambda= \left(\frac{t^{\beta \delta }}{h} \right)^{1+\frac{1}{\delta }}\Phi.
\end{equation}
The scaling form of the free energy, defined by (\ref{fhscaling2}) and
(\ref{lambdadef}), results in the finite-field Larkin Pikin equations 
\begin{eqnarray}\label{lpc}
\tilde{f}&=& h^{1+1/\delta} \Lambda\left(\frac{X}{h^{1/\beta /\delta}}
\right)+ \lambda\left(h^{\frac{1-\alpha }{\beta \delta }}
\Lambda'\left(\frac{X}{h^{1/\beta \delta }} \right) \right)^{2}\cr
t&=&X - 2 \lambda \left[\frac{1}{h^{(\alpha -1)/\beta \delta
}}\Lambda' (0)+ \frac{X}{ h^{\alpha /\beta \delta }}\Lambda'' (0) \right].
\end{eqnarray}
where details of the derivation of (\ref{lpc}) are presented in Appendix D.

We recall that criticality of $f[X,h]$ only occurs at
$X=0$ ($t=t_c$) indicating that $\tilde{f}[X,h]$ can only be critical at $X=0$. 
In the region of first order transitions $t[X]$ is non-monotonic with two 
points, a maxima and minima, where the gradient $dt/dX=0$ goes to zero.  
As we approach the critical field $h_{c}$, the maxima and minima merge 
together at a point of inflection, meeting at $X=0$.  As a result we 
deduce that the critical end point occurs when $X=0$ (for criticality) and at 
$dt/dX=0$ (merger of maximum and minimum).  More succinctly,
the critical endpoint  corresponds to the inflection
point in $t(X)$.  When we impose these two
conditions, we can solve for $h_{c}$ and $t_{c}=
(T_{CEP}-T^{(0)}_{c})/T^{(0)}_{c}$, which from  (\ref{lpc})
implies that 
\pagebreak
\begin{widetext}
\begin{eqnarray}\label{139}
h_{c}&=& (2 \lambda\Lambda'')^{\delta \beta /\alpha }, \qquad \qquad \qquad \qquad \quad\quad\quad\quad\quad(dt/dX = 0)\cr
t_{c}&=& - \frac{2\lambda \Lambda'}{h_{c}^{\frac{\alpha -1}{\delta \beta}}} = - h_{c}^{\frac{1}{\delta \beta }}\frac{\Lambda'}{\Lambda''}=
- (2 \lambda \Lambda'')^{\frac{1}{\alpha
}}\frac{\Lambda'}{\Lambda''}, \quad\qquad (X = 0)\end{eqnarray}
We can be sure that these quantities are both positive, because
$\Delta S = -\partial f[t,h]/\partial t= h^{\frac{1-\alpha }{\beta \delta }}\Lambda'$ is the change in
the entropy due to the field, and we expect this to be negative, so that $\Lambda'<0$
and hence $t_{c}>0$. Similarly, $-\frac{\partial^{2}f}{\partial
t^{2}}\sim \frac{\Delta C}{T}\sim h^{-\alpha /\delta \beta
}\Lambda''$. This quantity gets bigger as the field is reduced, so
that $\Lambda''>0$, guaranteeing that $h_{c}>0$ is real and positive.

\subsection{Field Behavior Close to the Quantum Critical Endpoint}

When we look at this problem as part of the approach to a QCP, we must
now include the amplitude $A_{I} (T_{c})= T_{c}^{(\alpha
-\tilde{\alpha }/\tilde{\Psi })}$.  The tuning parameter $t$ of the
classical calculation now becomes 
$\gf= g-g_{c} (T_{c})= g - u T_{c}^{\frac{1}{\Psi}}$. 
If we now expand around the finite
temperature critical point at a specific $T_{c}$, the singular
free energy of the clamped system is
\begin{equation}\label{}
f[\gf,h] = -
 h^{\frac{2-{\alpha}}{{\beta} {\delta} }} 
\Lambda\left(\frac{\delta g}{h^{1/\beta\delta }} \right)A_{I} (T_{c}).
\end{equation}

\noindent The Larkin Pikin equations now become
\begin{eqnarray}\label{qlpc}
\tilde{f}[X,h]&=& 
 -h^{\frac{2-{\alpha}}{{\beta} {\delta} }} 
A_{I} (T_{c})\Lambda\left(\frac{X}{h^{1/\beta \delta}}
\right)+ \lambda\left(h^{\frac{1-\alpha }{\beta \delta }}
A_{I} (T_{c})\Lambda'\left(\frac{X}{h^{1/\beta \delta }} \right) \right)^{2}\cr
\gf[X]&=&X - 2 \lambda A_{I} (T_{c}) \left[\frac{1}{h^{(\alpha -1)/\beta \delta
}}\Lambda' (0)+ \frac{X}{ h^{\alpha /\beta \delta }}\Lambda'' (0) \right].
\end{eqnarray}
where again $\tilde{f}[X,h]$ refers to the free energy of the unclamped system.
\end{widetext}

As discussed in the last section, the critical end point occurs at
$X=0$.
The critical end point is then at $g = -uT_{c}^{1/\tilde\Psi}+\gf[0]$.
When we do the subsequent algebra, we see that we get similar
equations to those we obtained at finite temperature (\ref{lpc})
with the replacement $\lambda\rightarrow \lambda A_{I} (T_{c})$.
Taking equations \eqref{139} and replacing $t_{c}\rightarrow \gf_c$
and $\lambda\rightarrow \lambda A_{I} (T_{c})$, we obtain
\begin{eqnarray}\label{l}
h_{c}&\propto & (\lambda A_{I} (T_{c}))^{\delta \beta /\alpha } =
(\lambda )^{\delta \beta /\alpha } (T_{c})^{\frac{\delta \beta ({\alpha}-\tilde\alpha
)}{\alpha\tilde{\Psi } }
} , \cr
\gf_{c}&\propto &  ( \lambda A_{I} (T_{c}) )^{\frac{1}{\alpha
}} = \lambda^{\frac{1}{\alpha}} (T_{c})^{\frac{(\alpha -\tilde{\alpha })}{\alpha \tilde{\Psi }}}.
\end{eqnarray}
These equations are valid in the plane of constant $T_{c}$. For small
$\lambda$ we can transform these expressions into the plane of
constant $g$, writing $T_{c}= (g_{c} (0) -g)^{\tilde{\Psi }}$, while the location of
the critical end point is at a  temperture  $T_{EP}=T_{c}+\delta T_{EP}$,
where $\delta T_{EP} = \gf (dT_{c}/dg)\propto \gf (g_{c}
(0)-g)^{\tilde{\Psi }-1}$, which then gives 
\begin{eqnarray}\label{ql}
h_{c}&\propto & \lambda^{\frac{\delta \beta}{\alpha }}
(g_{c} (0) -g)^{
\frac{\delta \beta ({\alpha}-\tilde\alpha
)}{\alpha }
}
\cr
\delta T_{EP}&\sim &  
\lambda^{\frac{1}{\alpha }}(g_{c} (0) -g)^{\frac{{\alpha }-\tilde\alpha }{{\alpha }}- (1-\tilde\Psi )}
\end{eqnarray}
where we have restored $g_{c} (0)$
For the Gaussian fixed point considered by Larkin and Pikin, 
with  $\tilde\alpha =0$, ${\alpha }=1/2$, ${\beta }=1/4$
${\delta }=5$, $\tilde{\Psi}  = 1/2$, we have
\begin{eqnarray}\label{l}
T_{c}&\sim& (g_{c}-g)^{1/2}\cr
h_{c}&\sim& \lambda^{5/2}(g_{c}-g)^{5/4}\sim T_{c}^{5/2}\cr
\delta T_{EP}& \sim & \lambda^{2}(g_{c}-g)^{1/2}
\end{eqnarray}
which yields a pointed, $V$-shaped ``anteater's tongue'' as
the surface of first-order transitions in the LP problem. In
Figure 7 we present a schematic of the evolution of
the critical endpoints as a function of $T_c$, and we note
that the full Larkin-Pikin phase diagram is displayed in Figure 1. 

%\figwidth=10cm
%\fg{anteater2.jpg}{}{Anteater's Tongue}

\color{black}
\figwidth=1.05\columnwidth
\fg{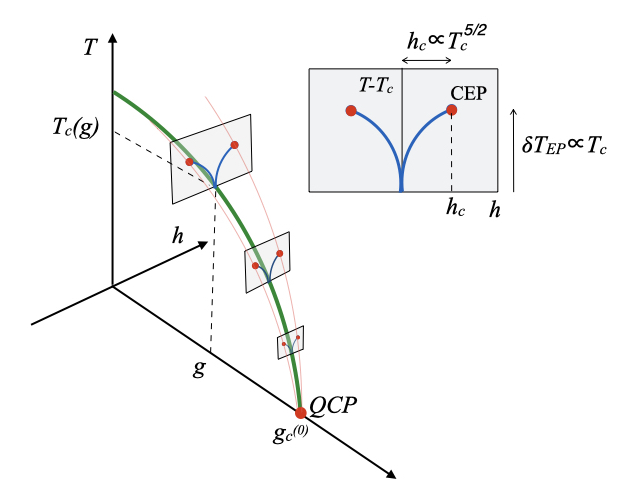}{Fig7}{Evolution of the critical end-points with $T_{c}$.}

\section{Implications for Observable Properties}

The specific heat exponent $\alpha$ of the clamped (fixed volume) system plays a key role in the universality of the classical Larkin-Pikin criterion (\ref{LP}) since 
the coupling of the order parameter to the lattice is a strain-energy density. For the scalar ($n=1$) case considered 
here, $\alpha > 0$, so that $\Delta \kappa$ is singular and the
finite-temperature transition is always first-order in the unclamped
(fixed pressure) system.  
By contrast for $d + z > 4$, the system is above its upper critical
dimension and there is a continuous transition at $T=0$ and
quantum annealed criticality.
The amplitudes of thermodynamic quantities will decrease with
temperature in the approach to the quantum critical point,
and we have specifically presented this behavior for the latent work
and the entropy.

A key motivation for our study has been recent
low-temperature experiments on polar insulators
that display quantum criticality even though their
classical transitions are first-order.
Many ferroelectrics have scalar order parameters with 
dynamical exponent $z=1$, so such three-dimensional
materials are in their 
marginal dimension;
logarithmic corrections to the bulk modulus are certainly 
present but they are not 
expected to be singular.
Indeed such contributions to the dielectric 
susceptibility, $\chi$, 
in the approach to ferroelectric 
quantum critical points have not been observed to 
date \cite{Rowley14}; furthermore here the temperature-dependence of 
$\chi$ is described well by a self-consistent 
Gaussian approach appropriate above its upper critical dimension \cite{Rowley14,Chandra17}. Therefore there may be a very
weak first-order quantum phase transition \cite{Das13} but 
experimentally 
it appears to be indistinguishable from 
a continuous one.  We note that near quantum criticality the main effect of 
long-range dipolar interactions, not included in this treatment, 
is to produce a gap in the logitudinal fluctuations, but the 
transverse fluctuations remain critical \cite{Rechester71,Khmelnitskii73,Roussev03}; the excellent agreement between theory and experiment at ferroelectric quantum criticality confirms that this is the case \cite{Rowley14,Chandra17}.  

Dielectric loss and hysteresis measurements can be 
used to probe the line of classical 
first-order transitions, and to determine the nature of the 
quantum phase transition.  The Gruneisen ratio ($\Gamma$), the 
ratio of the thermal expansion and the specific heat, is known to change signs across the quantum phase 
transition \cite{Zhu03,Garst05}; furthermore it is predicted to diverge at a 
3D ferroelectric quantum critical point as $\Gamma \propto \frac{1}{T^2}$ 
so this would be a good indicator 
of underlying quantum criticality \cite{Chandra17}.
Both the bulk modulus and the longitudinal sound velocity should display 
features near quantum annealed criticality, where specifics are 
material-dependent. 
Elastic anisotropy 
may drive this system
into an inhomogeneous state\cite{Bergman76,Bruno80,deMoura76}.  
The coupling of domain
dynamics to anisotropic strain has been studied 
classically for ferroelectrics \cite{Brierley14}, and implications for
the quantum case are a topic for future work.

\section{Discussion  and Open Questions}

In summary, we have developed a theoretical framework to describe compressible 
insulating systems that have classical first-order transitions 
and display pressure-induced quantum criticality.  We have generalized 
the Larkin-Pikin approach \cite{Larkin69b}  
to the quantum case using crossover scaling forms that describe both its classical and its quantum behavior.
We show that when the system is above its upper critical dimensionality, there is no latent work at
the quantum transition indicating that it is continuous.  We then include a field conjugate to the order
parameter, and derive the Larkin-Pikin phase diagram with three critical points, two classical and one quantum.
 Following the original Larkin-Pikin analysis, ours has been performed for a scalar order parameter and isotropic elasticity where the phase transition is first-order for all finite temperatures; here we show that for $d+z >4$ the
quantum transition is continuous.  The key point  is that a compressible material can host a quantum critical phase 
even if it displays a first-order transition at ambient pressure.  More generally the 
order of the classical phase transition can be different from its
quantum counterpart. 

An interesting question arising from our work, is whether the
Larkin Pikin mechanism can be understood in a broader field-theoretic
context. The long-range interaction that drives the Larkin Pikin mechanism
relies on the presence of a finite shear modulus: a rigidity of a
solid that is absent in a liquid. 
The Larkin Pikin derivation has a topological 
flavor, in that the $q=0$ ``boundary component'' of the strain which drives the 
long-range interaction, when integrated
around a closed loop on a torus is a topological invariant that counts 
the number of enclosed defects \eqref{topoloop} and which is closely
connected with the concept of torsion\cite{Katanaev2005}. 
In a system with boundaries, we still expect the long-range
interaction, but now derived from the boundary waves of the material.
There is thus a kind of bulk-boundary correspondence in the phenomenon
that may be topological in character.  One possibility here, is that
the Larkin Pikin interaction, which breaks the Lorentz invariance of the
short-range physics, is a kind of symmetry breaking anomaly\cite{Witten2016}.

Recently the possibility of a line of discontinuous transitions ending in a quantum critical point has also been
studied in frustrated spin models\cite{Kellermann19,Schmidt20}, in multiferroics, \cite{Narayan19,Chandra19} 
and in transition metal difluorides.\cite{Hancock15} 
There are also experiments on metallic 
systems \cite{Schmiedeshoff11,Tokawa13,Steppke13} that 
suggest quantum annealed criticality, so a quantum
generalization of the electronic case \cite{Pikin70}  
with possible links to previous work on metallic magnets should be 
pursued \cite{Brando16}; implications for doped paraelectric materials and polar metals \cite{Chandra17} will also be explored. Extension of this work to
quantum transitions between two distinct ordered states separated by first-order classical transitions may be relevant to the iron-based
superconductors \cite{Quader14} and to the enigmatic heavy 
fermion material 
$URu_2Si_2$ where quantum critical endpoints have been suggested
\cite{Chandra13}.

Finally the possibility of quantum annealed criticality in 
compressible materials, 
magnetic and ferroelectric, provides a new setting for the exploration
of exotic quantum phases where a broad temperature range 
can be probed with easily accessible pressures
due to the lattice-sensitivity of these systems. 
In particular, the elimination of the Larkin Pikin mechanism at $T=0$
exposes a bare quantum critical point,  a state of matter with quantum
fluctuations on all scales, with the potential for instabilities into
novel quantum phases. 

We have benefitted from discussions with colleagues including
A.V. Balatsky, L.B. Ioffe, D. Khmelnitskii, P.B. Littlewood and P.A. Volkov. PC
and PC  gratefully acknowledge the Centro Brasileiro de Pesquisas 
Fisicas (CBPF), Trinity College (Cambridge) and the Cavendish Laboratory
where this project was initiated.  
Early stages of this work were supported by 
CAPES and FAPERJ grants CAPES-AUXPE-EAE-705/2013 (PC and PC), 
FAPERJ-E-26/110.030/2103 (PC and PC), and
NSF grant DMR-1830707 (P. Coleman);
this work was also supported by 
grant DE-SC0020353 (P.Chandra) funded by the U.S.
Department of Energy, Office of Science.
MAC acknowledges the Brazilian agencies CNPq and FAPERJ for partial
financial support.
GGL acknowledges support from grant no. EP/K012984/1 of the ERPSRC and the CNPq/Science without Borders Program.
PC, PC and GGL thank the Aspen Center for Physics and NSF grant PHYS-1066293 
for hospitality where this work was further developed and discussed.
PC and PC thank S. Nakatsuji and the Institute for Solid State Physics
(U. Tokyo) for hospitality when he early stages of this work were
underway. 

\vskip0.2truein

\appendix

\begin{widetext}

\section{Gaussian Strain integration in the Classical case}

We would like to integrate out the Gaussian
elastic degrees of freedom from the action so that
the partition function takes the form
\begin{equation}\label{summary2a}
Z = 
\int {\cal D}[\psi] \int {\cal D}[u ] \ e^{- \mathscr{S}[\psi,u]} 
\longrightarrow 
Z = 
\int {\cal D}[\psi]e^{- S[\psi]}.  
\end{equation}
where the effective action S is a function of the order
parameter $\psi$.
We write 
\begin{equation}\label{Splusa}
S[\psi]= S_{L}[\psi ]+ \Delta S[\psi ] 
\end{equation}
where 
\begin{equation}\label{SLa}
S_{L}[\psi] = 
\frac{1}{T} \int d^3 x \left [ \frac{1}{2} (\partial_{\mu}\psi
)^{2}+ \frac{a}{2}\psi^{2}+ \frac{b}{4!}\psi^{4} \right]
\end{equation}
describes the physics of the order parameter in the clamped
system with tuning parameter $a \propto \frac{T-T_c}{T_C}$ and
$b > 0 $ as in (\ref{action0}) and (\ref{l3}).
Our task is to calculate the Gassian integral(\ref{deltaS}),
\begin{equation}\label{deltaS3}
e^{-\Delta S[\psi] }  = \int {\cal D}[e_{ab},u_{q}]e^{- ({\cal S}_{E}+{\cal S}_{I})}
\end{equation}
with
\begin{equation}\label{SBSI}
\mathscr{S}_{E}+\mathscr{S}_{I} = 
\frac{1}{T} \int d^{3}x \left[  \frac{1}{2}
\left(K-\frac{2}{3}\mu \right)
e_{ll}^{2} (\vec{x}) + \mu e_{ab} (\vec{x})^{2}
+ \lambda \psi^{2} (\vec{x}) e_{ll} (\vec{x}) \right].
\end{equation}
%where we have denoted $\sigma_{ab}= -{P}\delta_{ab}$ in terms
%of the pressure $P$. 
As discussed in the main text, we separate the strain field into 
its $q=0$ and finite $q$ components (\ref{splity}), 
\begin{equation}\label{}
e_{ab} (\vec{x}) = e_{ab} + \frac{1}{\sqrt{V}} \sum_{\vec{q}\neq 0} \frac{i}{2}
\left(q_{a}u_{b} (\vec{q}) + q_{b}u_{a} (\vec{q}) \right)e^{ i \vec{q}\cdot \vec{x}},
\end{equation}
so that the action (\ref{SBSI}) divides into two terms, 
$\mathscr{S}_E + \mathscr{S}_I = S[e_{ab},\psi ]+ S[u,\psi ]$. 
We next define the integrals
\begin{equation*}\label{S1}
\int d e_{ab} \ e^{-S[e_{ab},\psi ]} = e^{-S_{1}[\psi ]},
\end{equation*}
and 
\begin{equation}\label{S2}
\int {\cal D}[u]e^{-S[u,\psi ]} = e^{-S_{2}[\psi]}.
\end{equation}
that treat the $q=0$ and finite $q$ elastic contributions 
to (\ref{deltaS3}) respectively.

The uniform part of the action is
\begin{eqnarray}\label{seab}
S[e_{ab},\psi ] &=& 
\frac{V}{T} \left[  \frac{1}{2}
\left(K-\frac{2}{3}\mu \right)
e_{ll}^{2}  +\mu e_{ab}^{2}\right]
+ \frac{V}{T} \lambda \psi^{2}_{q=0} e_{ll} \cr
&=& \frac{1}{2} e_{ab}{\cal M}_{abcd} e_{cd} + v_{ab}e_{ab},
\end{eqnarray}
where
$\psi^{2}_{\vec{q}}= \frac{1}{V}\int d^{3}x\psi^{2} (\vec{x})e^{i \vec{q}\cdot
\vec{x}}
$
 is the Fourier transform of the fluctuations in ``energy
density'' and
\begin{eqnarray}\label{curlym}
{\cal M}_{abcd} &=& \frac{1}{T} \left\{K 
\overbrace {(\delta_{ab}\delta_{cd})}^{{\cal P}^{L}_{abcd}}+
2\mu
\overbrace {\left(\delta_{ac}\delta_{bd}-\frac{1}{3}\delta_{ab}\delta_{cd}
\right) }^{{\cal P}^{T}_{abcd}}\right\}, \\
v_{ab}&=& \frac{V}{T} \lambda \psi^{2}_{q=0} \delta_{ab}.
\end{eqnarray}
In (\ref{curlym}) ${\cal P}^L_{abcd}$ and ${\cal P}^T_{abcd}$ are independent
projection operators
(${\cal P}^{\Gamma}_{abef}{\cal P}^{\Gamma}_{efcd}={\cal P}^{\Gamma}_{abcd}$, $\Gamma\in L,
T$) associated wtih the longitudinal and transverse components of the
strain.

When we integrate over the uniform part of the strain field (\ref{seab}),
\begin{eqnarray}\label{uni1}
 S[e_{ab},\psi] = \frac{1}{2} e_{ab}{\cal M}_{abcd} e_{cd} + v_{ab}e_{ab}
\ \ \rightarrow \ \
S_{1}[\psi ]= - \frac{1}{2}v_{ab}{\cal M}^{-1}_{abcd}v_{cd}
\end{eqnarray}
Because of the independent nature of the projection 
operators ${\cal P}^{L,T}_{abcd}$ in (\ref{curlym}), we can write
the inverse of ${\cal M}$ as 
\begin{equation}\label{curlyminv}
{\cal M}^{-1}_{abcd} = \frac{T}{V}\left[\frac{1}{K} (\delta_{ab}\delta_{cd})+
\frac{1}{2\mu}\left(\delta_{ac}\delta_{bd}-\frac{1}{3}\delta_{ab}\delta_{cd}
\right) \right],
\end{equation}
so the Gaussian integral over the uniform part of the strain field
yields
\begin{equation}\label{result1}
S_{1}[\psi] = - \frac{1}{2}v_{ab}{\cal M}^{-1}_{abcd}v_{cd} =  
-\frac{V}{2T} \frac{\lambda^2}{K} (\psi^{2}_{q=0})^{2}.
\end{equation}
which can also be written as
\begin{equation}\label{uresult}
S_{1} [\psi] =  - \frac{\lambda^2}{2T}
\left(\frac{1}{K} \right)
\left[ \frac{1}{V} \int d^{3}x \ \int d^{3}x' \ \psi^{2}(\vec{x}) \ \psi^{2}(\vec{x}') \right].
\end{equation}

The nonuniform part of the action is  
\begin{equation}\label{}
S[u,\psi ] 
= 
 \frac{1}{T}
\sum_{\vec{q}\neq 0} \left( \frac{1}{2}
u^{*}_{a}({\vec{q}}) M_{ab}
u_{b} (\vec{q}) + \vec{a} (\vec{q})
\cdot \vec{u} (\vec{q})\right)
\end{equation}
where
\begin{eqnarray}\label{l}
M_{ab}&=& \left[ 
\left(K - \frac{2}{3}\mu \right)
q_{a}q_{b}
+ \mu \left(q^{2}\delta_{ab}+q_{a}q_{b} \right)
\right],\cr
\vec{a}_{q}&=& 
\left( 
 {i\lambda
}{\sqrt{V}}
\ \psi^{2}_{-q}\right) \vec{q}.
\end{eqnarray}
The matrix entering the fluctuating part of the action $S[u,\psi]$ 
in (\ref{S2}) can be 
projected into the longitudinal and transverse components of the strain 
\begin{equation}\label{}
M_{ab} (\vec{q}) = q^{2} \left[ 
\left(K  +\frac{4}{3}\mu \right)
\hat  q_{a}\hat q_{b}
+ \mu (\delta_{ab}- \hat  q_{a}\hat q_{b})
\right]
\end{equation}
where  $\hat q_{a}=q_{a}/q$ are the direction cosines of $\vec{q}$.  Inversion of this
matrix is then 
\begin{equation}\label{}
M^{-1}_{ab} (\vec{q}) = q^{-2} \left[ 
\left(K  +\frac{4}{3}\mu \right)^{-1}
\hat  q_{a}\hat q_{b}
+ \mu^{-1}(\delta_{ab}-\hat  q_{a}\hat q_{b})
\right],
\end{equation}
so the Gaussian integral over fluctuating part of the strain field
leads to
\begin{eqnarray}\label{l}
 S[u,\psi] = \frac{1}{T}
\sum_{\vec{q}\neq 0}\frac{1}{2}
u^{*}_{a}({\vec{q}}) M_{ab} (\vec{q})
u_{b} (\vec{q}) + \vec{a} (\vec{q})
\cdot \vec{u} (\vec{q})&\rightarrow& \cr
S_{2}[\psi ]&=& -\frac{1}{2T}\sum_{\vec{q}\neq  0} a_{a} (-\vec{q} )
M^{-1}_{ab} (\vec{q}) a_{b} (\vec{q})\cr
&=&-\frac{V}{2T} \sum_{\vec{q}\neq  0 }  \psi^{2}_{-q}\psi^{2}_{q}\frac{\lambda^{2}}{K+\frac{4}{3}\mu}
\end{eqnarray}
We can rewrite this as a sum over {\sl all} $\vec{q}$, plus a 
remainder at $\vec{q}=0$:
\begin{eqnarray}\label{result2}
S_{2}[\psi]
&=& 
-\frac{V}{2T} \sum_{\vec{q}}
\psi^{2}_{-q}\psi^{2}_{q}\frac{\lambda^{2}}{K+\frac{4}{3}\mu}
+\frac{V}{2T} (
\psi^{2}_{q=0})^{2}\frac{\lambda^{2}}{K+\frac{4}{3}\mu}\cr
&=& 
- \frac{1}{2T}\frac{\lambda^{2}}{K+\frac{4}{3}\mu}\int d^{3}x \psi^{4}
(\vec{x})
+\frac{V}{2T} (
\psi^{2}_{q=0})^{2}\frac{\lambda^{2}}{K+\frac{4}{3}\mu}
\end{eqnarray}
which can be reexpressed as 
\begin{equation}\label{qresult}
S_{2}[\psi] =
- \frac{\lambda^2}{2T}\left(\frac{1}{K+\frac{4}{3}\mu}\right)
\left\{\int d^{3}x \ \psi^{4}
(\vec{x})
- \left[ \frac{1}{V} \int d^{3}x \ \int d^{3}x' \ \psi^{2}(\vec{x}) \ \psi^{2}(\vec{x}') 
\right]\right\}.
\end{equation}
The first term is a local attraction while the second term
corresponds 
to a long-range repulsion. 

When we combine (\ref{uresult}) and (\ref{qresult}), we obtain
\begin{equation}\label{DeltaSsum}
\Delta S [\psi] =
- \frac{\lambda^2}{2T} \left\{\left(\frac{1}{K+\frac{4}{3}\mu}\right)
\int d^{3}x \ \psi^{4}
(\vec{x})
+
\left(\frac{1}{K} - \frac{1}{K+\frac{4}{3}\mu}\right)
\left[ \frac{1}{V} \int d^{3}x \ \int d^{3}x' \ \psi^{2}(\vec{x}) \ \psi^{2}(\vec{x}') 
\right]\right\}.
\end{equation}
Recalling (\ref{Splusa}), we note that we can 
group the first term  in (\ref{DeltaSsum}) in
the local $S_{L}[\psi]$ (\ref{SLa})  
to obtain the results 
\begin{equation}\label{aa22}
S [\psi] =  S_{L}[\psi,a,b^*] - \frac{\lambda^2}{2T}
\left(\frac{1}{K} - \frac{1}{K + \frac{4}{3}\mu}\right)
\left[ \frac{1}{V} \int d^{3}x \ \int d^{3}x' \ \psi^{2}(\vec{x}) \ \psi^{2}(\vec{x}') \right],
\end{equation} 
where 
\begin{align}\label{slocal2}
S_{L} [\psi,a,b^*]= \frac{1}{T}\int d^{3}x\left[
\frac{1}{2} (\partial_{\mu}\psi
)^{2}+ \frac{a}{2}\psi^{2}+ \frac{b^{*}}{4!}\psi^{4}
 \right]
\end{align}
with a renormalized local interaction
\begin{equation}\label{rbb2}
b^* = b - \frac{12\lambda^{2}}{K+\frac{4}{3}\mu}.
\end{equation}
as in the main text (equations (\ref{aa2}), (\ref{slocal}) and (\ref{rbb})).

\section{Gaussian Strain Integration in the Quantum Case}

We would like to integrate out the Gaussian
elastic degrees of freedom so that the partition function takes the
form 
\begin{equation}\label{summary2b}
Z = 
\int {\cal D}[\psi] \int {\cal D}[u ] \ e^{- \mathscr{S}[\psi,u]} 
\longrightarrow 
Z = 
\int {\cal D}[\psi]e^{- S[\psi]}.  
\end{equation}
where the integrals are over spacetime
(\ref{action0q}), 
$S = \int d^{4}x L\equiv \int_{0}^{\beta } d\tau
\int d^{3}x L$.
We write the effective action
\begin{equation}\label{Splusqb}
S[\psi]= S_{L}[\psi ]+ \Delta S[\psi ] 
\end{equation}
where
\begin{equation}\label{SLb}
S_{L}[\psi] = 
\int d^4 x \left [ \frac{1}{2} (\partial_{\mu}\psi
)^{2}+ \frac{a}{2}\psi^{2}+ \frac{b}{4!}\psi^{4} \right]
\end{equation} 
and 
\begin{equation}\label{deltaSqb}
e^{-\Delta S[\psi ]}=  \int {\cal D}[e,u]e^{- (\mathscr{S}_{E}[u] 
+\mathscr{S}_{I}[\psi,e])}
\end{equation}
with
\begin{equation}\label{SBSIq}
\mathscr{S}_{E}+ \mathscr{S}_{I} = 
\int d^{4}x \left[  
\frac{\rho }{2}\dot u_{l}^{2}+ \left(K-\frac{2}{3}\mu \right)
e_{ll}^{2} (\vec{x}) + \frac{1}{2}2\mu e_{ab} (\vec{x})^{2}
+ \lambda \psi^{2} (\vec{x}) e_{ll} (\vec{x}) \right].
\end{equation}
This action can be cast into matrix form
\begin{equation}\label{SBSIb}
\mathscr{S}_{E}+\mathscr{S}_{I}= \frac{1}{2} \sum_{q}
u_{i}M_{ij} u_{j} + \lambda \sum_{j}u_{j} \psi_{j}^{2} \rightarrow  \frac{\lambda^{2}}{2}
\sum_{i,j} \psi_{i}^{2}M^{-1}_{i,j}\psi_j^2.
\end{equation}
where now the summations
run over the discrete
wavevector and Matsubara frequencies $q \equiv (i\nu_n,\vec q)$, where 
$\nu_n=\frac{2\pi }{\beta  }n$, $\vec{ q} = \frac{2\pi}{L} (j,l,k)$.
As discussed in the main text, we separate out the static $\vec{ q}=0$ 
component of the strain tensor (\ref{splityQ1}), writing 
\begin{equation}\label{splityQ1b}
e_{ab} (x,\tau ) = e_{ab} + \frac{1}{\sqrt{V\beta} } \sum_{i\nu_n}
\sum_{\vec{q}\neq 0} \frac{i}{2}
\left(q_{a}u_{b} (q) + q_{b}u_{b} (q) \right)e^{ i (\vec{q}\cdot \vec{x}-\nu_n\tau)}.
\end{equation}
We note that there is no time-dependence in the uniform part of the
strain since we restrict ourselves to smooth Gaussian deformations of the solid
(see discussion in main text preceeding (\ref{splityQ1})). 
However the fluctuating component includes all Matsubara frequencies; 
with these caveats, the quantum integration of the strain fields 
closely follows that of the classical case.
Given the form of the elastic tensor (\ref{splityQ1b}), the action 
(\ref{SBSIb}) naturally
divides into two terms, 
\begin{equation}\label{Sdivqb}
\mathscr{S}_E + \mathscr{S}_I = S[e_{ab},\psi ]+S[u,\psi ]
\end{equation}
corresponding to the distinct unifom and finite $\vec{q}$ contributions
to the strain, and 
we define the respective integrals
\begin{equation*}
\int de_{ab} e^{-S[e_{ab},\psi ]} = e^{-S_{1}[\psi ]}
\end{equation*}
and 
\begin{equation}\label{}
\int {\cal D}[u]e^{-S[u,\psi ]} = e^{-S_{2}[\psi]}.
\end{equation}
so that 
\begin{equation}\label{deltaSdef}
\Delta S [\psi] = S_1[\psi] + S_2[\psi]. 
\end{equation}

The uniform part of the action
\begin{eqnarray}\label{ll1}
S[e_{ab},\psi ] &=& 
\int d^4x  \left[  \frac{1}{2}
\left(K-\frac{2}{3}\mu \right)
e_{ll}^{2}  + \frac{1}{2}2\mu e_{ab}^{2}\right]
+ \frac{V}{T} (\lambda \psi^{2}_{q=0}) e_{ll} \cr
&=& \frac{1}{2} e_{ab}{\cal M}_{abcd} e_{cd} + v_{ab}e_{ab},
\end{eqnarray}
where 
\begin{eqnarray}\label{curlymq}
{\cal M}_{abcd} &=& \left[K (\delta_{ab}\delta_{cd})+
2\mu\left(\delta_{ac}\delta_{bd}-\frac{1}{3}\delta_{ab}\delta_{cd}
\right) \right], \cr
v_{ab}&=& {V\beta } \lambda \psi^{2}_{q=0}\delta_{ab},
\end{eqnarray}
is similar to the classical case (\ref{curlym}), but now 
\begin{equation}\label{stFT}
\psi^{2}_{q}= \frac{1}{V\beta } \int d^{4}x \ \psi^{2} ({x})e^{-i (\vec{q}\cdot \vec{x}-\nu_n\tau)}
\end{equation}
is the spacetime Fourier transform of the order parameter intensity.
When we integrate over the uniform part of the strain field, we obtain
\begin{eqnarray}\label{uresultb}
 S[e_{ab},\psi] = \frac{1}{2} e_{ab}{\cal M}_{abcd} e_{cd} + v_{ab}e_{ab}
\ \ \rightarrow \ \
S_{1}[\psi ]= - \frac{1}{2}v_{ab}{\cal M}^{-1}_{abcd}v_{cd} 
\end{eqnarray}\label{s1resultb}
which, as in the classical case, can be reexpressed as
\begin{equation}\label{S1qq}
S_{1}[\psi] = -\frac{\lambda^2 \beta V}{2K} (\psi^{2}_{q=0})^{2}
\end{equation}
using 
(\ref{curlym}), (\ref{curlyminv}) and (\ref{curlymq}) where
$\beta = \frac{1}{T}$.

The nonuniform part of the elastic contribution to (\ref{deltaSqb}) is 
\begin{equation}\label{}
S[u,\psi ] 
= 
 \sum_{i\nu_n}\sum_{\vec{q}\neq 0}
\left(
\frac{1}{2}
u^{*}_{a}(q) M_{ab}
u_{b} (q) + \vec{a} (q)
\cdot \vec{u} (q) \right),
\end{equation}
where $q = (\vec{q},i\nu_n)$ and we use Roman letters (e.g. $a,b$) to
denote spatial variables so that $q_a$ is a spatial component of q.
Here
\begin{eqnarray}\label{l}
M_{ab}&=& \left[ \rho \nu_n^2
\left(K - \frac{2}{3}\mu \right)
q_{a}q_{b}
+ \mu \left(q^{2}\delta_{ab}+q_{a}q_{b} \right)
\right],\cr
\vec{a}_{q}&=& 
\left( 
 {i\lambda
}{\sqrt{V\beta }}
\ \psi^{2}_{-q}\right) \vec{q}.
\end{eqnarray}
This matrix can be 
projected into its longitudinal and transverse components
\begin{equation}\label{}
M_{ab}= \left[ 
\left(\rho \nu_n^2 + (K  +\frac{4}{3}\mu ) \right)
\hat  q_{a}\hat q_{b}
+ \left(\rho \nu_n^2 + \mu \right)(\delta_{ab}- \hat  q_{a}\hat q_{b})
\right],
\end{equation}
where  $\hat q_{a}=q_{a}/q$ is the unit vector. Inversion of this
matrix is then 
\begin{equation}\label{}
M^{-1}_{ab}=  \left[
\frac{1}{\rho (\nu_n^2+c_L^2 q^{2})} 
\hat  q_{a}\hat q_{b}
+ \frac{1}{\rho (\nu_n^2+c_T^2q^2)}
(
\delta_{ab}-\hat  q_{a}\hat q_{b})
\right],
\end{equation}
where 
\begin{equation}\label{cLb}
c_{L}^{2} = \frac{K+\frac{4}{3}\mu}{\rho }, \qquad c_{T}^{2} = \frac{2\mu}{\rho }
\end{equation}
are the longitudinal and transverse sound velocities; the two terms
appearing in $M^{-1}$ are recognized as the propagators for
longitudinal and tranverse phonons.

When we integrate over the fluctuating component of the strain field,
only the longitudinal phonons couple to the order parameter:  
\begin{eqnarray}\label{l}
\frac{1}{2}
\sum_{i\nu_n}\sum_{\vec{q}\neq 0}
u^{*}_{a}(q) M_{ab} (q)
u_{b} (q) + \vec{a} (q)
\cdot \vec{u} (q)&\rightarrow& \cr
S_{2}[\psi ]&=& -\frac{1}{2}\sum_{i\nu_n}\sum_{\vec{q}\neq  0} a_{a} (-q )
M^{-1}_{ab} (q) a_{b} (q)\cr
&=&
-\frac{V\beta \lambda^{2}}{2 } \sum_{i\nu_n,\vec{q}\neq  0 }
\psi^{2}_{-q}\psi^{2}_{q}
\left(\frac{q^{2}}{\rho \nu_n^2+ (K+\frac{4}{3}\mu) q^2} \right).
\end{eqnarray}
In this last term, 
\begin{equation}\label{}
\left(\frac{q^{2}}{\rho \nu_n^2+ (K+\frac{4}{3}\mu) q^2} \right)
\end{equation}
the $\vec{q}=0$ term vanishes for any finite 
$\nu_n$, but in the case where $\nu_n=0$, the limiting
$\vec{q}\rightarrow 0$ form of this term is finite:
\begin{equation}\label{}
\left. 
\left(\frac{q^{2}}{\rho \nu_n^2+ (K+\frac{4}{3}\mu) q^2} \right)\right\vert_{\vec{q}\rightarrow 0} =
\left\{\begin{array}{cc}
0 & \nu_{n}\neq  0\cr
\frac{1}{K+\frac{4}{3}\mu} & \nu_n=0.
\end{array}
 \right.
\end{equation}
We can thus replace
\begin{equation}\label{l}
\sum_{i\nu_n,\vec{q}\neq  0 }
\psi^{2}_{-q}\psi^{2}_{q}
\left(\frac{q^{2}}{\rho \nu_n^2+ (K+\frac{4}{3}\mu) q^2}
\right)
\rightarrow 
\sum_{i\nu_n,\vec{q}}
\psi^{2}_{-q}\psi^{2}_{q}
\left(
\frac{q^{2}}{\rho \nu_n^2+ (K+\frac{4}{3}\mu) q^2} 
\right)
-
 \frac{(\psi^{2}_{q=0})^{2}}{K+\frac{4}{3}\mu}.
\end{equation}
so that 
\begin{equation}\label{S2qb1}
S_{2}[\psi ] = 
\frac{V\beta \lambda^{2}}{2 (K+\frac{4}{3}\mu)} (\psi^{2}_{q=0})^{2} - 
\frac{V\beta \lambda^{2}}{2}\sum_{i\nu_n,\vec{q}}
\psi^{2}_{-q}\psi^{2}_{q}
\left(
\frac{q^{2}}{\rho \nu_n^2+ (K+\frac{4}{3}\mu) q^2} 
\right).
\end{equation}
which can be rewritten as
\begin{equation}\label{S2qq}
S_{2}[\psi ] = 
\frac{V\beta \lambda^{2}}{2 (K+\frac{4}{3}\mu)} \left \{(\psi^{2}_{q=0})^{2} - 
\sum_{i\nu_n,\vec{q}}
\psi^{2}_{-q}\psi^{2}_{q}
\left(1 - 
\frac  {\nu_n^2/c_L^2 } 
{\nu_n^2/c_L^2+ q^2}
\right)\right\}
\end{equation}
where $c_L$ is defined in (\ref{cLb}).

If we now combine (\ref{S1qq}) and (\ref{S2qq}), 
recalling (\ref{deltaSdef}),
we obtain 
\begin{equation}\label{S12qq}
\Delta S = 
-\frac{V\beta \lambda^{2}}{2} 
\left \{ \left( \frac{1}{K} - \frac{1}{K+\frac{4}{3}\mu} \right)
(\psi^{2}_{q=0})^{2} 
- 
- \left(\frac{1}{K+\frac{4}{3}\mu}\right)
\sum_{i\nu_n,\vec{q}}
\psi^{2}_{-q}\psi^{2}_{q}
\left(1 - 
\frac  {\nu_n^2/c_L^2 } 
{\nu_n^2/c_L^2+ q^2}
\right)\right\}
\end{equation}
The useful spacetime expression
\begin{equation}
\int d^4x \frac{e^{i(q - q')x}}{\beta V} = \delta_{qq'}
\end{equation}
allows us to rewrite (\ref{S12qq}) in spacetime coordinates as
\begin{equation}\label{S12x}
\Delta S = 
-\frac{\lambda^2}{2} 
\left \{ \frac{1}{\beta V \kappa}
\int d^4x \  d^4x' \psi^{2}({x}) \psi^2({x}') 
- \left(\frac{1}{K+\frac{4}{3}\mu}\right)
\int d^4x \ [\psi^4({x})]
+ \int d^4x \ d^4x' \psi^2({x}) V_{dyn} (x - x') \psi^2({x}') \right\} 
\end{equation}
where
\begin{equation}
V_{dyn} (x - x') = \frac{1}{\beta V} \sum_{\vec{q},i\nu_n} 
\frac 
{e^{-i (\vec{q}\cdot\vec{x}-\nu_n\tau)}}
{{(K + \frac{4}{3}\mu)}}
\frac  {\nu_n^2/c_L^2 } {\nu_n^2/c_L^2+ q^2}
\end{equation}
with
\begin{equation}\label{}
\frac{1}{\kappa} = \frac{1}{K}- \frac{1}{K+\frac{4}{3}\mu}
\end{equation}
is the effective Bulk modulus.
Recalling (\ref{Splusqb}), we note that we can 
group the second and third terms  in (\ref{S12x}) in
$S_{L}[\psi]$ (\ref{SLb})  
to obtain the results 
\begin{equation}\label{bresultq1}
S [\psi] =  S_{L}[\psi] - \frac{\lambda^2}{2}
\left(\frac{1}{K} - \frac{1}{K + \frac{4}{3}\mu}\right)
\left[ \frac{1}{\beta V} \int d^{4}x \ \int d^{4}x' \ \psi^{2}({x}) \ \psi^{2}({x}') \right],
\end{equation} 
with
\begin{align}\label{slocal2}
S_{L} [\psi]= \int d^{4}x\left[
\frac{1}{2} (\partial_{\mu}\psi
)^{2}+ \frac{a}{2}\psi^{2}+ \frac{b^{*}}{4!}\psi^{4}
+ \frac{1}{2} \psi^2({x}) V_{dyn}(x - x') \psi^2({x}')
 \right]
\end{align}
where the Fourier transform of $V_{dyn}(x-x')$ is
\begin{equation}\label{vphononb}
V_{dyn} (q)= 
\frac{\lambda^{2}}{K+\frac{4}{3}\mu}\left(\frac{\nu^{2}_n/c_{L}^{2}}{\vec{q}^{2}+\nu^{2}_n/c_{L}^{2}}
\right),
\end{equation}
and
\begin{equation}\label{rbb2b}
b^* = b - \frac{12\lambda^{2}}{K+\frac{4}{3}\mu}.
\end{equation}
as in the main text (equations (\ref{a2q}), (\ref{slocalq}) and (\ref{rbbq2})).
We note that (\ref{vphononb}) is finite and 
non-singular in the limit $q \rightarrow 0$, 
and has the same scaling dimensions as the original local 
repulsive interaction in (\ref{SLb}); from a scaling perspective it is thus
local and will not afect the critical behavior of the clamped system.

\section{Derivation of the Amplitude Factors for the Crossover Scaling}

In order to generalize the Larkin-Pikin argument to $T \rightarrow 0$,
we need to introduce a crossover scaling form for the clamped
free energy $f$ in (\ref{fxq}) and (\ref{axq}) that is applicable near both
the classical and the quantum critical points.\cite{Continentino17}
At a finite temperature, the location of the phase transition  is
shifted by the thermal fluctuations, so that 
\begin{equation}\label{a1c}
g_{c} (T) = g_{c} (0) - uT^{\frac{1}{\tilde\Psi }}.
\end{equation}
where $\tilde\Psi $ is called the shift exponent; 
we note that if the effective
dimension of the quantum system is below its upper
critical dimension $\tilde\Psi = \tilde\nu z$. 
For convenience, we'll take the zero temperature QCP critical coupling
constant to be zero, $g_{c} (0) =0$.  Now temperature is a finite
size correction to the quantum critical point,  and the 
free energy is determined by a crossover function 
\begin{equation}\label{a2c}
f (g,T)= g^{2-\tilde\alpha} \Phi \left(\frac{T^{\frac{1}{\tilde\Psi }}}{g}\right).
\end{equation}
that describes both the quantum critical
point, and the finite temperature classical critical point of the
clamped system (see Figure 5); here we will use the convention that $\alpha$ 
and $\tilde{\alpha}$ refer to the classical and quantum exponents respectively.  
At a finite temperature the critical
behavior now occurs at the shifted value of $g_{c} (T)$, governed by
the finite temperature specific heat exponent ${\alpha }$.
For a fixed $T$ scan with small $g-g_{c} (T)$, the singular part of
the free energy is  
\begin{equation}\label{a3}
f (g,T)= (g-g_{c} (T))^{2-{\alpha }} A_{I} (T).
\end{equation}
where $A_I(T)$ is the amplitude factor for the classical critical
point occuring at $T_{c}=T$.  

The scaling form (\ref{a2c}) allows us to determine the form of this
amplitude factor.  The crucial observation is that the classical
critical point occurs at a value $T^{1/\Psi }/g = -1/u$, so that $\Phi
(\vec{x})$ must have a singularity of the form $(g+uT^{\frac{1}{\tilde\Psi
}})^{2-\alpha }= g^{2-\alpha } (1+ u  \frac{T^{1/\tilde\Psi
}}{g})^{2-{\alpha }}\sim (1+u x)^{2- \alpha }$, so that the scaling function
takes the form
\begin{equation}\label{a0}
{\Phi (\vec{x})} = 
{(1+ux)^{2-{\alpha }}}
\tilde{ \Phi } (\vec{x}),
\end{equation}
 To see this in detail, 
let us rewrite (\ref{a2c}) as 
\begin{eqnarray}\label{l}
f (g,T)&=& 
(g-g_{c} (T))^{2-{\alpha }}\frac{g^{2-\tilde\alpha
}}{(g-g_{c} (T))^{2-{\alpha }}}  \Phi \left(\frac{T^{\frac{1}{\Psi }}}{g}\right)
\cr 
&=& 
(g-g_{c} (T))^{2-{\alpha }}\frac{ g^{{\alpha }-\tilde\alpha
}}{(1+ u 
\frac{T^{\frac{1}{\tilde\Psi }}}{g}
)^{2-{\alpha }}}  \Phi
\left(\frac{T^{\frac{1}{\tilde\Psi }}}{g}\right).
\end{eqnarray}
In other words,
\begin{equation}\label{}
f (g,T) = 
(g-g_{c} (T))^{2-{\alpha }}
g^{{\alpha }-\tilde\alpha}
\tilde{\Phi}
\left(\frac{T^{\frac{1}{\Psi }}}{g}\right), 
\end{equation}
where 
\begin{equation}\label{}
\tilde{ \Phi } (\vec{x})= \frac{\Phi (\vec{x})}{(1+ux)^{2-{\alpha }}}.
\end{equation}
To assure a classical phase
transition at finite temperature with the right exponent, 
the cross-over function $\tilde{\Phi }$ must be smooth around 
$x=-1/u$, in otherwords, the original cross-over function
contains a hidden singularity at $x=-1/u$ and 
factorizes as follows:
\begin{equation}\label{}
{\Phi (\vec{x})} = 
{(1+ux)^{2-{\alpha }}}
\tilde{ \Phi } (\vec{x}),
\end{equation}
as inferred in (\ref{a0}).
Thus at finite temperature,
the singularity at zero temperature splits into a shifted
singularity with modified exponent $2-{\alpha }$ 
\begin{equation}\label{}
f (g,T)=(g-g_{c} (T))^{2-{\alpha }} A[g,T],
\end{equation}
where the amplitude factor is given by 
\begin{eqnarray}\label{l}
A[g,T]=  g^{{\alpha }-\tilde\alpha}
\tilde{\Phi}
\left(\frac{T^{\frac{1}{\Psi }}}{g}\right).
 \end{eqnarray}

Suppose we carry out a sweep at constant temperature $T_{c}$
(Fig. \ref{fig4})
, then 
near the classical critical line, we may replace $g=g_{c} (T_{c})=
uT_{c}^{\frac{1}{\tilde\Psi }}$ , so that
$T_{c}^{1/\Psi}/g=-1/u$, 
inside the cross-over function and
\begin{equation}\label{xover1}
f[g,T_{c}]\sim 
(g-g_{c} )^{2-{\alpha }} A_{I} (T_{c})
\end{equation}
where 
\begin{equation}\label{xover1b}
A_{I}(T_{c})= A[g_{c} (T),T_{c}]= 
a_{1}T_{c}^{\left(\frac{{\alpha }-\tilde\alpha }{\tilde\Psi } \right)},
\end{equation}
and $a_{1}=\left[ 
u^{{\alpha }-\tilde\alpha }
\tilde{\Phi}
\left(-\frac{1}{u}\right)\right]
$.
Likewise, if we carry out a sweep through the phase transition at
constant $g_{c}$ (Fig. \ref{fig4}), 
then we can write
\begin{equation}\label{xover2a}
f[g_{c},T]\sim (T-T_{c})^{2-{\alpha }}A_{II} (g_{c})
\end{equation}
where 
\begin{eqnarray}\label{xover2b}
A_{II} (g_{c})&=& \left(
\frac{dg_{c}}{dT} \right)^{2-{\alpha }}g_{c}^{{\alpha
}-\tilde\alpha }\tilde{\Phi } \left(-\frac{1}{u}\right)
\cr
&=& a_{2}g_{c}^{(1-\Psi ) (2-{\alpha })+ ({\alpha}-\tilde\alpha )}.
\end{eqnarray}
with $a_{2}=\left(\frac{u^{\Psi }}{\tilde\Psi } \right)^{2-{\alpha }}\tilde{\Phi } \left(-\frac{1}{u}\right)$.
Summarizing the amplitude factors are then
\begin{eqnarray}\label{l}
A_{I} (T_{c})&=& a_{1}T_{c}^{\left(\frac{{\alpha }-\tilde\alpha }{\tilde\Psi
} \right)},\cr\cr
A_{II} (g_{c})&=& a_{2}g_{c}^{(1-\Psi ) (2-{\alpha })+ ({\alpha}-\tilde\alpha )}.
\end{eqnarray}
as given in \eqref{ampfactors} in the main text.

\section{Derivation of the Larkin-Pikin Equations in Finite
Field}

We use our scaling form for the free energy
\begin{equation}\label{fhscalingd}
f\propto - t^{2-\alpha } \Phi\left(\frac{h}{t^{\beta \delta }} \right).
\end{equation}
where
$h$ is the dimensionless external field,  
$t= (T-T^0_{c})/T^0_{c}$ is the reduced temperature and
$T_c^0$ is the transition temperature of the clamped system.
We want to describe the behavior of (\ref{fhscalingd}) when
$t$ is small compared with $h^{1/\beta \delta }$.  
In this limit we know that $f\propto h^{\frac{1}{\delta }+1}$ and 
(\ref{fhscalingd}) 
can be rewritten
\begin{equation}\label{fhscaling2d}
f\propto - h^{\frac{1}{\delta }+1}\Lambda\left(\frac{t}{h^{1/\beta
\delta }} \right)= 
-
 h^{\frac{2-\alpha}{\beta \delta }} 
\Lambda\left(\frac{t}{h^{1/\beta\delta }} \right)
\end{equation}
where,using the identity $2-\alpha = \beta (1+\delta )$, 
we have substituted $(\delta +1)/\delta  = (2-\alpha )/\delta \beta $,
Comparing (\ref{fhscalingd}) and (\ref{fhscaling2d}), we obtain
\begin{equation}\label{}
\Lambda= \left(\frac{t^{\beta \delta }}{h} \right)^{1+\frac{1}{\delta }}\Phi.
\end{equation}
If  $y= \frac{t}{h^{1/\beta\delta}}$ and 
$z = y^{-\beta \delta }$, then we have
\begin{equation}\label{}
\Lambda (y) = y^{\beta (1+\delta )}\Phi (y^{-\beta \delta })= y^{2-\alpha }\Phi (y^{-\beta \delta }).
\end{equation}

We note that at large values of 
$z =h/t^{\beta \delta }$, small values of $y= \frac{t}{h^{1/\beta
\delta }}$, the free energy can be expanded perturbatively in $y$ around
$y=0$, so that
\begin{equation}\label{fcxh}
f[X,h] = - h^{\frac{2-\alpha}{\beta \delta }} 
\left[\Lambda (0)+ \frac{X}{h^{1/\beta \delta
}}\Lambda' (0)+ \frac{X^{2}}{2 h^{2/\beta \delta }}\Lambda'' (0) \right].
\end{equation}
where we have replaced $t$ by the parameterized variable $X$ of the
unclamped material.
This means that
\begin{equation}\label{fcpxh}
\frac{\partial f}{\partial X} \equiv f'_{X}= - \left[\frac{1}{h^{(\alpha -1)/\beta \delta
}}\Lambda' (0)+ \frac{X}{ h^{\alpha /\beta \delta }}\Lambda'' (0) \right].
\end{equation}
When (\ref{fcxh}) and (\ref{fcpxh}) are input into
the LP equations (\ref{fx}) and (\ref{ax}),
The Larkin Pikin equations in a finite field become
\begin{eqnarray}\label{lpcd}
\tilde{f}&=& h^{1+1/\delta} \Lambda\left(\frac{X}{h^{1/\beta /\delta}}
\right)+ \lambda\left(h^{\frac{1-\alpha }{\beta \delta }}
\Lambda'\left(\frac{X}{h^{1/\beta \delta }} \right) \right)^{2}\cr
t&=&X - 2 \lambda \left[\frac{1}{h^{(\alpha -1)/\beta \delta
}}\Lambda' (0)+ \frac{X}{ h^{\alpha /\beta \delta }}\Lambda'' (0) \right].
\end{eqnarray}
which are exactly the equations (\ref{lpc}) in the main text.

\end{widetext}

%\bibliography{QAC}

\begin{thebibliography}{58}%
\makeatletter
\providecommand \@ifxundefined [1]{%
 \@ifx{#1\undefined}
}%
\providecommand \@ifnum [1]{%
 \ifnum #1\expandafter \@firstoftwo
 \else \expandafter \@secondoftwo
 \fi
}%
\providecommand \@ifx [1]{%
 \ifx #1\expandafter \@firstoftwo
 \else \expandafter \@secondoftwo
 \fi
}%
\providecommand \natexlab [1]{#1}%
\providecommand \enquote  [1]{``#1''}%
\providecommand \bibnamefont  [1]{#1}%
\providecommand \bibfnamefont [1]{#1}%
\providecommand \citenamefont [1]{#1}%
\providecommand \href@noop [0]{\@secondoftwo}%
\providecommand \href [0]{\begingroup \@sanitize@url \@href}%
\providecommand \@href[1]{\@@startlink{#1}\@@href}%
\providecommand \@@href[1]{\endgroup#1\@@endlink}%
\providecommand \@sanitize@url [0]{\catcode `\\12\catcode `\$12\catcode
  `\&12\catcode `\#12\catcode `\^12\catcode `\_12\catcode `\%12\relax}%
\providecommand \@@startlink[1]{}%
\providecommand \@@endlink[0]{}%
\providecommand \url  [0]{\begingroup\@sanitize@url \@url }%
\providecommand \@url [1]{\endgroup\@href {#1}{\urlprefix }}%
\providecommand \urlprefix  [0]{URL }%
\providecommand \Eprint [0]{\href }%
\providecommand \doibase [0]{http://dx.doi.org/}%
\providecommand \selectlanguage [0]{\@gobble}%
\providecommand \bibinfo  [0]{\@secondoftwo}%
\providecommand \bibfield  [0]{\@secondoftwo}%
\providecommand \translation [1]{[#1]}%
\providecommand \BibitemOpen [0]{}%
\providecommand \bibitemStop [0]{}%
\providecommand \bibitemNoStop [0]{.\EOS\space}%
\providecommand \EOS [0]{\spacefactor3000\relax}%
\providecommand \BibitemShut  [1]{\csname bibitem#1\endcsname}%
\let\auto@bib@innerbib\@empty
%</preamble>
\bibitem [{\citenamefont {Larkin}\ and\ \citenamefont
  {Pikin}(1969)}]{Larkin69b}%
  \BibitemOpen
  \bibfield  {author} {\bibinfo {author} {\bibfnamefont {A.}~\bibnamefont
  {Larkin}}\ and\ \bibinfo {author} {\bibfnamefont {S.}~\bibnamefont {Pikin}},\
  }\href {http://www.jetp.ac.ru/cgi-bin/dn/e_029_05_0891.pdf} {\bibfield
  {journal} {\bibinfo  {journal} {Sov. Phys. JETP}\ }\textbf {\bibinfo {volume}
  {29}},\ \bibinfo {pages} {891} (\bibinfo {year} {1969})}\BibitemShut
  {NoStop}%
\bibitem [{\citenamefont {Belitz}\ \emph {et~al.}(1999)\citenamefont {Belitz},
  \citenamefont {Kirkpatrick},\ and\ \citenamefont {Vojta}}]{Belitz99}%
  \BibitemOpen
  \bibfield  {author} {\bibinfo {author} {\bibfnamefont {D.}~\bibnamefont
  {Belitz}}, \bibinfo {author} {\bibfnamefont {T.~R.}\ \bibnamefont
  {Kirkpatrick}}, \ and\ \bibinfo {author} {\bibfnamefont {T.}~\bibnamefont
  {Vojta}},\ }\href {\doibase 10.1103/PhysRevLett.82.4707} {\bibfield
  {journal} {\bibinfo  {journal} {Phys. Rev. Lett.}\ }\textbf {\bibinfo
  {volume} {82}},\ \bibinfo {pages} {4707} (\bibinfo {year}
  {1999})}\BibitemShut {NoStop}%
\bibitem [{\citenamefont {Grigera}\ \emph {et~al.}(2001)\citenamefont
  {Grigera}, \citenamefont {Perry}, \citenamefont {Schofield}, \citenamefont
  {Chiao}, \citenamefont {Julian}, \citenamefont {Lonzarich}, \citenamefont
  {Kieda}, \citenamefont {Meano}, \citenamefont {Millis},\ and\ \citenamefont
  {Mackenzie}}]{Grigera01}%
  \BibitemOpen
  \bibfield  {author} {\bibinfo {author} {\bibfnamefont {S.}~\bibnamefont
  {Grigera}}, \bibinfo {author} {\bibfnamefont {R.}~\bibnamefont {Perry}},
  \bibinfo {author} {\bibfnamefont {A.}~\bibnamefont {Schofield}}, \bibinfo
  {author} {\bibfnamefont {M.}~\bibnamefont {Chiao}}, \bibinfo {author}
  {\bibfnamefont {S.}~\bibnamefont {Julian}}, \bibinfo {author} {\bibfnamefont
  {G.}~\bibnamefont {Lonzarich}}, \bibinfo {author} {\bibfnamefont
  {S.}~\bibnamefont {Kieda}}, \bibinfo {author} {\bibfnamefont
  {Y.}~\bibnamefont {Meano}}, \bibinfo {author} {\bibfnamefont
  {A.}~\bibnamefont {Millis}}, \ and\ \bibinfo {author} {\bibfnamefont
  {A.}~\bibnamefont {Mackenzie}},\ }\href
  {http://science.sciencemag.org/content/294/5541/329} {\bibfield  {journal}
  {\bibinfo  {journal} {Science}\ }\textbf {\bibinfo {volume} {294}},\ \bibinfo
  {pages} {329} (\bibinfo {year} {2001})}\BibitemShut {NoStop}%
\bibitem [{\citenamefont {Chubukov}\ \emph {et~al.}(2004)\citenamefont
  {Chubukov}, \citenamefont {P\'epin},\ and\ \citenamefont
  {Rech}}]{Chubukov04}%
  \BibitemOpen
  \bibfield  {author} {\bibinfo {author} {\bibfnamefont {A.~V.}\ \bibnamefont
  {Chubukov}}, \bibinfo {author} {\bibfnamefont {C.}~\bibnamefont {P\'epin}}, \
  and\ \bibinfo {author} {\bibfnamefont {J.}~\bibnamefont {Rech}},\ }\href
  {\doibase 10.1103/PhysRevLett.92.147003} {\bibfield  {journal} {\bibinfo
  {journal} {Phys. Rev. Lett.}\ }\textbf {\bibinfo {volume} {92}},\ \bibinfo
  {pages} {147003} (\bibinfo {year} {2004})}\BibitemShut {NoStop}%
\bibitem [{\citenamefont {Belitz}\ \emph {et~al.}(2005)\citenamefont {Belitz},
  \citenamefont {Kirkpatrick},\ and\ \citenamefont {Vojta}}]{Belitz05}%
  \BibitemOpen
  \bibfield  {author} {\bibinfo {author} {\bibfnamefont {D.}~\bibnamefont
  {Belitz}}, \bibinfo {author} {\bibfnamefont {T.~R.}\ \bibnamefont
  {Kirkpatrick}}, \ and\ \bibinfo {author} {\bibfnamefont {T.}~\bibnamefont
  {Vojta}},\ }\href {\doibase 10.1103/RevModPhys.77.579} {\bibfield  {journal}
  {\bibinfo  {journal} {Rev. Mod. Phys.}\ }\textbf {\bibinfo {volume} {77}},\
  \bibinfo {pages} {579} (\bibinfo {year} {2005})}\BibitemShut {NoStop}%
\bibitem [{\citenamefont {Maslov}\ \emph {et~al.}(2006)\citenamefont {Maslov},
  \citenamefont {Chubukov},\ and\ \citenamefont {Saha}}]{Maslov06}%
  \BibitemOpen
  \bibfield  {author} {\bibinfo {author} {\bibfnamefont {D.~L.}\ \bibnamefont
  {Maslov}}, \bibinfo {author} {\bibfnamefont {A.~V.}\ \bibnamefont
  {Chubukov}}, \ and\ \bibinfo {author} {\bibfnamefont {R.}~\bibnamefont
  {Saha}},\ }\href {\doibase 10.1103/PhysRevB.74.220402} {\bibfield  {journal}
  {\bibinfo  {journal} {Phys. Rev. B}\ }\textbf {\bibinfo {volume} {74}},\
  \bibinfo {pages} {220402} (\bibinfo {year} {2006})}\BibitemShut {NoStop}%
\bibitem [{\citenamefont {Rech}\ \emph {et~al.}(2006)\citenamefont {Rech},
  \citenamefont {P\'epin},\ and\ \citenamefont {Chubukov}}]{Rech06}%
  \BibitemOpen
  \bibfield  {author} {\bibinfo {author} {\bibfnamefont {J.}~\bibnamefont
  {Rech}}, \bibinfo {author} {\bibfnamefont {C.}~\bibnamefont {P\'epin}}, \
  and\ \bibinfo {author} {\bibfnamefont {A.~V.}\ \bibnamefont {Chubukov}},\
  }\href {\doibase 10.1103/PhysRevB.74.195126} {\bibfield  {journal} {\bibinfo
  {journal} {Phys. Rev. B}\ }\textbf {\bibinfo {volume} {74}},\ \bibinfo
  {pages} {195126} (\bibinfo {year} {2006})}\BibitemShut {NoStop}%
\bibitem [{\citenamefont {Kirkpatrick}\ and\ \citenamefont
  {Belitz}(2012)}]{Kirkpatrick12}%
  \BibitemOpen
  \bibfield  {author} {\bibinfo {author} {\bibfnamefont {T.~R.}\ \bibnamefont
  {Kirkpatrick}}\ and\ \bibinfo {author} {\bibfnamefont {D.}~\bibnamefont
  {Belitz}},\ }\href {\doibase 10.1103/PhysRevB.85.134451} {\bibfield
  {journal} {\bibinfo  {journal} {Phys. Rev. B}\ }\textbf {\bibinfo {volume}
  {85}},\ \bibinfo {pages} {134451} (\bibinfo {year} {2012})}\BibitemShut
  {NoStop}%
\bibitem [{\citenamefont {Brando}\ \emph {et~al.}(2016)\citenamefont {Brando},
  \citenamefont {Belitz}, \citenamefont {Grosche},\ and\ \citenamefont
  {Kirkpatrick}}]{Brando16}%
  \BibitemOpen
  \bibfield  {author} {\bibinfo {author} {\bibfnamefont {M.}~\bibnamefont
  {Brando}}, \bibinfo {author} {\bibfnamefont {D.}~\bibnamefont {Belitz}},
  \bibinfo {author} {\bibfnamefont {F.~M.}\ \bibnamefont {Grosche}}, \ and\
  \bibinfo {author} {\bibfnamefont {T.~R.}\ \bibnamefont {Kirkpatrick}},\
  }\href {\doibase 10.1103/RevModPhys.88.025006} {\bibfield  {journal}
  {\bibinfo  {journal} {Rev. Mod. Phys.}\ }\textbf {\bibinfo {volume} {88}},\
  \bibinfo {pages} {025006} (\bibinfo {year} {2016})}\BibitemShut {NoStop}%
\bibitem [{\citenamefont {Chandra}\ \emph {et~al.}(1990)\citenamefont
  {Chandra}, \citenamefont {Coleman},\ and\ \citenamefont
  {Larkin}}]{Chandra90}%
  \BibitemOpen
  \bibfield  {author} {\bibinfo {author} {\bibfnamefont {P.}~\bibnamefont
  {Chandra}}, \bibinfo {author} {\bibfnamefont {P.}~\bibnamefont {Coleman}}, \
  and\ \bibinfo {author} {\bibfnamefont {A.}~\bibnamefont {Larkin}},\ }\href
  {https://link.aps.org/doi/10.1103/PhysRevLett.64.88} {\bibfield  {journal}
  {\bibinfo  {journal} {Phys. Rev. Lett.}\ }\textbf {\bibinfo {volume} {64}},\
  \bibinfo {pages} {88} (\bibinfo {year} {1990})}\BibitemShut {NoStop}%
\bibitem [{\citenamefont {Chandra}\ and\ \citenamefont
  {Coleman}(1995)}]{Chandra95b}%
  \BibitemOpen
  \bibfield  {author} {\bibinfo {author} {\bibfnamefont {P.}~\bibnamefont
  {Chandra}}\ and\ \bibinfo {author} {\bibfnamefont {P.}~\bibnamefont
  {Coleman}},\ }in\ \href@noop {} {\emph {\bibinfo {booktitle} {Les Houches
  Lecture Notes (Session LVI)}}},\ \bibinfo {editor} {edited by\ \bibinfo
  {editor} {\bibfnamefont {B.}~\bibnamefont {Doucot}}\ and\ \bibinfo {editor}
  {\bibfnamefont {J.}~\bibnamefont {Zinn-Justin}}}\ (\bibinfo  {publisher}
  {Elsevier},\ \bibinfo {address} {Amsterdam},\ \bibinfo {year} {1995})\ pp.\
  \bibinfo {pages} {495--594}\BibitemShut {NoStop}%
\bibitem [{\citenamefont {Balents}(2010)}]{Balents10}%
  \BibitemOpen
  \bibfield  {author} {\bibinfo {author} {\bibfnamefont {L.}~\bibnamefont
  {Balents}},\ }\href {https://www.nature.com/articles/nature08917} {\bibfield
  {journal} {\bibinfo  {journal} {Nature}\ }\textbf {\bibinfo {volume} {464}},\
  \bibinfo {pages} {199} (\bibinfo {year} {2010})}\BibitemShut {NoStop}%
\bibitem [{\citenamefont {Zacharias}\ \emph {et~al.}(2015)\citenamefont
  {Zacharias}, \citenamefont {Rosch},\ and\ \citenamefont
  {Garst}}]{Zacharias15}%
  \BibitemOpen
  \bibfield  {author} {\bibinfo {author} {\bibfnamefont {M.}~\bibnamefont
  {Zacharias}}, \bibinfo {author} {\bibfnamefont {A.}~\bibnamefont {Rosch}}, \
  and\ \bibinfo {author} {\bibfnamefont {M.}~\bibnamefont {Garst}},\
  }\href@noop {} {\bibfield  {journal} {\bibinfo  {journal} {Eur. Phys. J.
  Special Topics}\ }\textbf {\bibinfo {volume} {224}},\ \bibinfo {pages} {1021}
  (\bibinfo {year} {2015})}\BibitemShut {NoStop}%
\bibitem [{\citenamefont {Norman}(2016)}]{Norman16}%
  \BibitemOpen
  \bibfield  {author} {\bibinfo {author} {\bibfnamefont {M.~R.}\ \bibnamefont
  {Norman}},\ }\href {\doibase 10.1103/RevModPhys.88.041002} {\bibfield
  {journal} {\bibinfo  {journal} {Rev. Mod. Phys.}\ }\textbf {\bibinfo {volume}
  {88}},\ \bibinfo {pages} {041002} (\bibinfo {year} {2016})}\BibitemShut
  {NoStop}%
\bibitem [{\citenamefont {Paul}\ and\ \citenamefont {Garst}(2017)}]{Paul17}%
  \BibitemOpen
  \bibfield  {author} {\bibinfo {author} {\bibfnamefont {I.}~\bibnamefont
  {Paul}}\ and\ \bibinfo {author} {\bibfnamefont {M.}~\bibnamefont {Garst}},\
  }\href {\doibase 10.1103/PhysRevLett.118.227601} {\bibfield  {journal}
  {\bibinfo  {journal} {Phys. Rev. Lett.}\ }\textbf {\bibinfo {volume} {118}},\
  \bibinfo {pages} {227601} (\bibinfo {year} {2017})}\BibitemShut {NoStop}%
\bibitem [{\citenamefont {Fernandes}\ \emph {et~al.}(2018)\citenamefont
  {Fernandes}, \citenamefont {Orth},\ and\ \citenamefont
  {Schmalian}}]{Fernandes18}%
  \BibitemOpen
  \bibfield  {author} {\bibinfo {author} {\bibfnamefont {R.~M.}\ \bibnamefont
  {Fernandes}}, \bibinfo {author} {\bibfnamefont {P.~P.}\ \bibnamefont {Orth}},
  \ and\ \bibinfo {author} {\bibfnamefont {J.}~\bibnamefont {Schmalian}},\
  }\href {https://arxiv.org/abs/1804.00818} {\bibfield  {journal} {\bibinfo
  {journal} {arXiv:1804.00818}\ } (\bibinfo {year} {2018})}\BibitemShut
  {NoStop}%
\bibitem [{\citenamefont {Ishidate}\ \emph {et~al.}(1997)\citenamefont
  {Ishidate}, \citenamefont {Abe}, \citenamefont {Takahashi},\ and\
  \citenamefont {M\^ori}}]{Ishidate97}%
  \BibitemOpen
  \bibfield  {author} {\bibinfo {author} {\bibfnamefont {T.}~\bibnamefont
  {Ishidate}}, \bibinfo {author} {\bibfnamefont {S.}~\bibnamefont {Abe}},
  \bibinfo {author} {\bibfnamefont {H.}~\bibnamefont {Takahashi}}, \ and\
  \bibinfo {author} {\bibfnamefont {N.}~\bibnamefont {M\^ori}},\ }\href
  {\doibase 10.1103/PhysRevLett.78.2397} {\bibfield  {journal} {\bibinfo
  {journal} {Phys. Rev. Lett.}\ }\textbf {\bibinfo {volume} {78}},\ \bibinfo
  {pages} {2397} (\bibinfo {year} {1997})}\BibitemShut {NoStop}%
\bibitem [{\citenamefont {Suski}\ \emph {et~al.}(1983)\citenamefont {Suski},
  \citenamefont {Takaoka}, \citenamefont {Murase},\ and\ \citenamefont
  {Porowski}}]{Suski83}%
  \BibitemOpen
  \bibfield  {author} {\bibinfo {author} {\bibfnamefont {T.}~\bibnamefont
  {Suski}}, \bibinfo {author} {\bibfnamefont {S.}~\bibnamefont {Takaoka}},
  \bibinfo {author} {\bibfnamefont {K.}~\bibnamefont {Murase}}, \ and\ \bibinfo
  {author} {\bibfnamefont {S.}~\bibnamefont {Porowski}},\ }\href
  {https://www.sciencedirect.com/science/article/pii/0038109883904763}
  {\bibfield  {journal} {\bibinfo  {journal} {Solid State Communications}\
  }\textbf {\bibinfo {volume} {45}},\ \bibinfo {pages} {259} (\bibinfo {year}
  {1983})}\BibitemShut {NoStop}%
\bibitem [{\citenamefont {Horiuchi}\ \emph {et~al.}(2015)\citenamefont
  {Horiuchi}, \citenamefont {Kobayashi}, \citenamefont {Kumai}, \citenamefont
  {Minami}, \citenamefont {Kagawa},\ and\ \citenamefont {Tokura}}]{Horiuchi15}%
  \BibitemOpen
  \bibfield  {author} {\bibinfo {author} {\bibfnamefont {S.}~\bibnamefont
  {Horiuchi}}, \bibinfo {author} {\bibfnamefont {K.}~\bibnamefont {Kobayashi}},
  \bibinfo {author} {\bibfnamefont {R.}~\bibnamefont {Kumai}}, \bibinfo
  {author} {\bibfnamefont {N.}~\bibnamefont {Minami}}, \bibinfo {author}
  {\bibfnamefont {F.}~\bibnamefont {Kagawa}}, \ and\ \bibinfo {author}
  {\bibfnamefont {Y.}~\bibnamefont {Tokura}},\ }\href
  {https://www.nature.com/articles/ncomms8469} {\bibfield  {journal} {\bibinfo
  {journal} {Nature Communications}\ }\textbf {\bibinfo {volume} {6}},\
  \bibinfo {pages} {7469} (\bibinfo {year} {2015})}\BibitemShut {NoStop}%
\bibitem [{\citenamefont {Rowley}\ \emph {et~al.}(2014)\citenamefont {Rowley},
  \citenamefont {Spalek}, \citenamefont {Smith}, \citenamefont {Dean},
  \citenamefont {Itoh}, \citenamefont {Scott}, \citenamefont {Lonzarich},\ and\
  \citenamefont {Saxena}}]{Rowley14}%
  \BibitemOpen
  \bibfield  {author} {\bibinfo {author} {\bibfnamefont {S.}~\bibnamefont
  {Rowley}}, \bibinfo {author} {\bibfnamefont {L.}~\bibnamefont {Spalek}},
  \bibinfo {author} {\bibfnamefont {R.}~\bibnamefont {Smith}}, \bibinfo
  {author} {\bibfnamefont {M.}~\bibnamefont {Dean}}, \bibinfo {author}
  {\bibfnamefont {M.}~\bibnamefont {Itoh}}, \bibinfo {author} {\bibfnamefont
  {J.}~\bibnamefont {Scott}}, \bibinfo {author} {\bibfnamefont
  {G.}~\bibnamefont {Lonzarich}}, \ and\ \bibinfo {author} {\bibfnamefont
  {S.}~\bibnamefont {Saxena}},\ }\href
  {https://www.nature.com/articles/nphys2924} {\bibfield  {journal} {\bibinfo
  {journal} {Nature Physics}\ }\textbf {\bibinfo {volume} {10}},\ \bibinfo
  {pages} {367} (\bibinfo {year} {2014})}\BibitemShut {NoStop}%
\bibitem [{\citenamefont {Chandra}\ \emph {et~al.}(2017)\citenamefont
  {Chandra}, \citenamefont {Lonzarich}, \citenamefont {Rowley},\ and\
  \citenamefont {Scott}}]{Chandra17}%
  \BibitemOpen
  \bibfield  {author} {\bibinfo {author} {\bibfnamefont {P.}~\bibnamefont
  {Chandra}}, \bibinfo {author} {\bibfnamefont {G.}~\bibnamefont {Lonzarich}},
  \bibinfo {author} {\bibfnamefont {S.}~\bibnamefont {Rowley}}, \ and\ \bibinfo
  {author} {\bibfnamefont {J.}~\bibnamefont {Scott}},\ }\href
  {http://iopscience.iop.org/article/10.1088/1361-6633/aa82d2} {\bibfield
  {journal} {\bibinfo  {journal} {Rep. Prog. Phys.}\ }\textbf {\bibinfo
  {volume} {80}},\ \bibinfo {pages} {112502} (\bibinfo {year}
  {2017})}\BibitemShut {NoStop}%
\bibitem [{\citenamefont {Landau}\ and\ \citenamefont
  {Lifshitz}(1986)}]{Landau86}%
  \BibitemOpen
  \bibfield  {author} {\bibinfo {author} {\bibfnamefont {L.}~\bibnamefont
  {Landau}}\ and\ \bibinfo {author} {\bibfnamefont {E.}~\bibnamefont
  {Lifshitz}},\ }\href@noop {} {\emph {\bibinfo {title} {Theory of Elasticity,
  3rd Edition}}}\ (\bibinfo  {publisher} {Pergamon Press},\ \bibinfo {year}
  {1986})\BibitemShut {NoStop}%
\bibitem [{\citenamefont {G.~Borchhardt}\ and\ \citenamefont
  {Rost}(1976)}]{Borchhardt76}%
  \BibitemOpen
  \bibfield  {author} {\bibinfo {author} {\bibfnamefont {G.~S.}\ \bibnamefont
  {G.~Borchhardt}}\ and\ \bibinfo {author} {\bibfnamefont {A.}~\bibnamefont
  {Rost}},\ }\href@noop {} {\bibfield  {journal} {\bibinfo  {journal} {Phys.
  Stat. Sol.}\ }\textbf {\bibinfo {volume} {88}},\ \bibinfo {pages} {143}
  (\bibinfo {year} {1976})}\BibitemShut {NoStop}%
\bibitem [{\citenamefont {Lines}\ and\ \citenamefont {Glass}(1977)}]{Lines77}%
  \BibitemOpen
  \bibfield  {author} {\bibinfo {author} {\bibfnamefont {M.}~\bibnamefont
  {Lines}}\ and\ \bibinfo {author} {\bibfnamefont {A.}~\bibnamefont {Glass}},\
  }\href@noop {} {\emph {\bibinfo {title} {{Principles and Applications of
  Ferroelectrics and Related Materials}}}}\ (\bibinfo  {publisher} {Clarendon
  Press, Oxford},\ \bibinfo {year} {1977})\BibitemShut {NoStop}%
\bibitem [{\citenamefont {{S}achdev}(1999)}]{sachdev_qpt_book}%
  \BibitemOpen
  \bibfield  {author} {\bibinfo {author} {\bibfnamefont {S.}~\bibnamefont
  {{S}achdev}},\ }\href@noop {} {\emph {\bibinfo {title} {{\sl Quantum Phase
  Transitions}}}}\ (\bibinfo  {publisher} {Cambridge University Press},\
  \bibinfo {address} {Cambridge, U.K.},\ \bibinfo {year} {1999})\BibitemShut
  {NoStop}%
\bibitem [{\citenamefont {Millis}(1993)}]{Millis1993}%
  \BibitemOpen
  \bibfield  {author} {\bibinfo {author} {\bibfnamefont {A.~J.}\ \bibnamefont
  {Millis}},\ }\href@noop {} {\bibfield  {journal} {\bibinfo  {journal}
  {Physical review B, Condensed matter}\ }\textbf {\bibinfo {volume} {48}},\
  \bibinfo {pages} {7183} (\bibinfo {year} {1993})}\BibitemShut {NoStop}%
\bibitem [{\citenamefont {Continentino}(2017)}]{Continentino17}%
  \BibitemOpen
  \bibfield  {author} {\bibinfo {author} {\bibfnamefont {M.~A.}\ \bibnamefont
  {Continentino}},\ }\href@noop {} {\emph {\bibinfo {title} {{Quantum Scaling
  in Many-Body Systems}}}}\ (\bibinfo  {publisher} {Cambridge Unversity
  Press},\ \bibinfo {year} {2017})\BibitemShut {NoStop}%
\bibitem [{\citenamefont {Gehring}(2008)}]{Gehring08}%
  \BibitemOpen
  \bibfield  {author} {\bibinfo {author} {\bibfnamefont {G.}~\bibnamefont
  {Gehring}},\ }\href
  {http://iopscience.iop.org/article/10.1209/0295-5075/82/60004/meta}
  {\bibfield  {journal} {\bibinfo  {journal} {Europhys. Lett.}\ }\textbf
  {\bibinfo {volume} {82}},\ \bibinfo {pages} {60004} (\bibinfo {year}
  {2008})}\BibitemShut {NoStop}%
\bibitem [{\citenamefont {Gehring}\ and\ \citenamefont
  {Ahmed}(2010)}]{Gehring10}%
  \BibitemOpen
  \bibfield  {author} {\bibinfo {author} {\bibfnamefont {G.}~\bibnamefont
  {Gehring}}\ and\ \bibinfo {author} {\bibfnamefont {A.}~\bibnamefont
  {Ahmed}},\ }\href {https://aip.scitation.org/doi/10.1063/1.3366612}
  {\bibfield  {journal} {\bibinfo  {journal} {J. Appl. Phys.}\ }\textbf
  {\bibinfo {volume} {107}},\ \bibinfo {pages} {09E125} (\bibinfo {year}
  {2010})}\BibitemShut {NoStop}%
\bibitem [{\citenamefont {Mineev}(2017)}]{Mineev17}%
  \BibitemOpen
  \bibfield  {author} {\bibinfo {author} {\bibfnamefont {V.~P.}\ \bibnamefont
  {Mineev}},\ }\href
  {http://iopscience.iop.org/article/10.3367/UFNe.2016.04.037771} {\bibfield
  {journal} {\bibinfo  {journal} {Phys.-Usp.}\ }\textbf {\bibinfo {volume}
  {60}},\ \bibinfo {pages} {121} (\bibinfo {year} {2017})}\BibitemShut
  {NoStop}%
\bibitem [{\citenamefont {Bean}\ and\ \citenamefont {Rodbell}(1962)}]{Bean62}%
  \BibitemOpen
  \bibfield  {author} {\bibinfo {author} {\bibfnamefont {C.}~\bibnamefont
  {Bean}}\ and\ \bibinfo {author} {\bibfnamefont {D.}~\bibnamefont {Rodbell}},\
  }\href {https://link.aps.org/doi/10.1103/PhysRev.126.104} {\bibfield
  {journal} {\bibinfo  {journal} {Phys. Rev.}\ }\textbf {\bibinfo {volume}
  {126}},\ \bibinfo {pages} {104} (\bibinfo {year} {1962})}\BibitemShut
  {NoStop}%
\bibitem [{\citenamefont {Bergman}\ and\ \citenamefont
  {Halperin}(1976)}]{Bergman76}%
  \BibitemOpen
  \bibfield  {author} {\bibinfo {author} {\bibfnamefont {D.}~\bibnamefont
  {Bergman}}\ and\ \bibinfo {author} {\bibfnamefont {B.}~\bibnamefont
  {Halperin}},\ }\href {https://link.aps.org/doi/10.1103/PhysRevB.13.2145}
  {\bibfield  {journal} {\bibinfo  {journal} {Phys. Rev. B}\ }\textbf {\bibinfo
  {volume} {13}},\ \bibinfo {pages} {2145} (\bibinfo {year}
  {1976})}\BibitemShut {NoStop}%
\bibitem [{\citenamefont {Sak}(1974)}]{Sak74}%
  \BibitemOpen
  \bibfield  {author} {\bibinfo {author} {\bibfnamefont {J.}~\bibnamefont
  {Sak}},\ }\href@noop {} {\bibfield  {journal} {\bibinfo  {journal} {Phys.
  Rev. B}\ }\textbf {\bibinfo {volume} {10}},\ \bibinfo {pages} {3957}
  (\bibinfo {year} {1974})}\BibitemShut {NoStop}%
\bibitem [{\citenamefont {Bruno}\ and\ \citenamefont {Sak}(1980)}]{Bruno80}%
  \BibitemOpen
  \bibfield  {author} {\bibinfo {author} {\bibfnamefont {J.}~\bibnamefont
  {Bruno}}\ and\ \bibinfo {author} {\bibfnamefont {J.}~\bibnamefont {Sak}},\
  }\href {https://link.aps.org/doi/10.1103/PhysRevB.22.3302} {\bibfield
  {journal} {\bibinfo  {journal} {Phys. Rev. B}\ }\textbf {\bibinfo {volume}
  {22}},\ \bibinfo {pages} {3302} (\bibinfo {year} {1980})}\BibitemShut
  {NoStop}%
\bibitem [{\citenamefont {Aharony}(1976)}]{Aharony76}%
  \BibitemOpen
  \bibfield  {author} {\bibinfo {author} {\bibfnamefont {A.}~\bibnamefont
  {Aharony}},\ }in\ \href@noop {} {\emph {\bibinfo {booktitle} {Phase
  Transitions and Critical Phenemenon}}},\ \bibinfo {editor} {edited by\
  \bibinfo {editor} {\bibfnamefont {C.}~\bibnamefont {Domb}}\ and\ \bibinfo
  {editor} {\bibfnamefont {M.}~\bibnamefont {Green}}}\ (\bibinfo  {publisher}
  {Academic Press},\ \bibinfo {year} {1976})\BibitemShut {NoStop}%
\bibitem [{\citenamefont {Vieira}\ and\ \citenamefont
  {Carvalho}(2014)}]{Vieira2014}%
  \BibitemOpen
  \bibfield  {author} {\bibinfo {author} {\bibfnamefont {W.}~\bibnamefont
  {Vieira}}\ and\ \bibinfo {author} {\bibfnamefont {P.}~\bibnamefont
  {Carvalho}},\ }\href@noop {} {\bibfield  {journal} {\bibinfo  {journal} {Eur.
  Phys. Lett.}\ }\textbf {\bibinfo {volume} {108}},\ \bibinfo {pages} {21001}
  (\bibinfo {year} {2014})}\BibitemShut {NoStop}%
\bibitem [{\citenamefont {Pfeuty}\ \emph {et~al.}(1974)\citenamefont {Pfeuty},
  \citenamefont {Jasnow},\ and\ \citenamefont {Fisher}}]{pfeutyfisher74}%
  \BibitemOpen
  \bibfield  {author} {\bibinfo {author} {\bibfnamefont {P.}~\bibnamefont
  {Pfeuty}}, \bibinfo {author} {\bibfnamefont {D.}~\bibnamefont {Jasnow}}, \
  and\ \bibinfo {author} {\bibfnamefont {M.~E.}\ \bibnamefont {Fisher}},\
  }\href {\doibase 10.1103/PhysRevB.10.2088} {\bibfield  {journal} {\bibinfo
  {journal} {Phys. Rev. B}\ }\textbf {\bibinfo {volume} {10}},\ \bibinfo
  {pages} {2088} (\bibinfo {year} {1974})}\BibitemShut {NoStop}%
\bibitem [{\citenamefont {Das}(2013)}]{Das13}%
  \BibitemOpen
  \bibfield  {author} {\bibinfo {author} {\bibfnamefont {N.}~\bibnamefont
  {Das}},\ }\href
  {https://www.worldscientific.com/doi/pdf/10.1142/S0217979213500288}
  {\bibfield  {journal} {\bibinfo  {journal} {Int. J. Mod. Phys. B.}\ }\textbf
  {\bibinfo {volume} {27}},\ \bibinfo {pages} {1350028} (\bibinfo {year}
  {2013})}\BibitemShut {NoStop}%
\bibitem [{\citenamefont {Rechester}(1971)}]{Rechester71}%
  \BibitemOpen
  \bibfield  {author} {\bibinfo {author} {\bibfnamefont {A.}~\bibnamefont
  {Rechester}},\ }\href {http://jetp.ac.ru/cgi-bin/dn/e_033_02_0423.pdf}
  {\bibfield  {journal} {\bibinfo  {journal} {Sov. Phys. JETP}\ }\textbf
  {\bibinfo {volume} {33}},\ \bibinfo {pages} {423} (\bibinfo {year}
  {1971})}\BibitemShut {NoStop}%
\bibitem [{\citenamefont {Khmelnitskii}\ and\ \citenamefont
  {Shneerson}(1973)}]{Khmelnitskii73}%
  \BibitemOpen
  \bibfield  {author} {\bibinfo {author} {\bibfnamefont {D.}~\bibnamefont
  {Khmelnitskii}}\ and\ \bibinfo {author} {\bibfnamefont {V.}~\bibnamefont
  {Shneerson}},\ }\href {http://jetp.ac.ru/cgi-bin/dn/e_037_01_0164.pdf}
  {\bibfield  {journal} {\bibinfo  {journal} {Sov. Phys. JETP}\ }\textbf
  {\bibinfo {volume} {37}},\ \bibinfo {pages} {164} (\bibinfo {year}
  {1973})}\BibitemShut {NoStop}%
\bibitem [{\citenamefont {Roussev}\ and\ \citenamefont
  {Millis}(2003)}]{Roussev03}%
  \BibitemOpen
  \bibfield  {author} {\bibinfo {author} {\bibfnamefont {R.}~\bibnamefont
  {Roussev}}\ and\ \bibinfo {author} {\bibfnamefont {A.}~\bibnamefont
  {Millis}},\ }\href {https://link.aps.org/doi/10.1103/PhysRevB.67.014105}
  {\bibfield  {journal} {\bibinfo  {journal} {Physical Review B}\ }\textbf
  {\bibinfo {volume} {67}},\ \bibinfo {pages} {014105} (\bibinfo {year}
  {2003})}\BibitemShut {NoStop}%
\bibitem [{\citenamefont {Zhu}\ \emph {et~al.}(2003)\citenamefont {Zhu},
  \citenamefont {Garst}, \citenamefont {Rosch},\ and\ \citenamefont
  {Si}}]{Zhu03}%
  \BibitemOpen
  \bibfield  {author} {\bibinfo {author} {\bibfnamefont {L.}~\bibnamefont
  {Zhu}}, \bibinfo {author} {\bibfnamefont {M.}~\bibnamefont {Garst}}, \bibinfo
  {author} {\bibfnamefont {A.}~\bibnamefont {Rosch}}, \ and\ \bibinfo {author}
  {\bibfnamefont {Q.}~\bibnamefont {Si}},\ }\href
  {https://link.aps.org/doi/10.1103/PhysRevLett.91.066404} {\bibfield
  {journal} {\bibinfo  {journal} {Physical Review Letters}\ }\textbf {\bibinfo
  {volume} {91}},\ \bibinfo {pages} {066404} (\bibinfo {year}
  {2003})}\BibitemShut {NoStop}%
\bibitem [{\citenamefont {Garst}\ and\ \citenamefont {Rosch}(2005)}]{Garst05}%
  \BibitemOpen
  \bibfield  {author} {\bibinfo {author} {\bibfnamefont {M.}~\bibnamefont
  {Garst}}\ and\ \bibinfo {author} {\bibfnamefont {A.}~\bibnamefont {Rosch}},\
  }\href {https://link.aps.org/doi/10.1103/PhysRevB.72.205129} {\bibfield
  {journal} {\bibinfo  {journal} {Physical Review B}\ }\textbf {\bibinfo
  {volume} {72}},\ \bibinfo {pages} {205129} (\bibinfo {year}
  {2005})}\BibitemShut {NoStop}%
\bibitem [{\citenamefont {de~Moura}\ \emph {et~al.}(1976)\citenamefont
  {de~Moura}, \citenamefont {Lubensky}, \citenamefont {Imry},\ and\
  \citenamefont {Aharony}}]{deMoura76}%
  \BibitemOpen
  \bibfield  {author} {\bibinfo {author} {\bibfnamefont {M.}~\bibnamefont
  {de~Moura}}, \bibinfo {author} {\bibfnamefont {T.}~\bibnamefont {Lubensky}},
  \bibinfo {author} {\bibfnamefont {Y.}~\bibnamefont {Imry}}, \ and\ \bibinfo
  {author} {\bibfnamefont {A.}~\bibnamefont {Aharony}},\ }\href
  {https://link.aps.org/doi/10.1103/PhysRevB.13.2176} {\bibfield  {journal}
  {\bibinfo  {journal} {Phys. Rev. B}\ }\textbf {\bibinfo {volume} {13}},\
  \bibinfo {pages} {2176} (\bibinfo {year} {1976})}\BibitemShut {NoStop}%
\bibitem [{\citenamefont {Brierley}\ and\ \citenamefont
  {Littlewood}(2014)}]{Brierley14}%
  \BibitemOpen
  \bibfield  {author} {\bibinfo {author} {\bibfnamefont {R.~T.}\ \bibnamefont
  {Brierley}}\ and\ \bibinfo {author} {\bibfnamefont {P.~B.}\ \bibnamefont
  {Littlewood}},\ }\href {\doibase 10.1103/PhysRevB.89.184104} {\bibfield
  {journal} {\bibinfo  {journal} {Phys. Rev. B}\ }\textbf {\bibinfo {volume}
  {89}},\ \bibinfo {pages} {184104} (\bibinfo {year} {2014})}\BibitemShut
  {NoStop}%
\bibitem [{\citenamefont {Katanaev}(2005)}]{Katanaev2005}%
  \BibitemOpen
  \bibfield  {author} {\bibinfo {author} {\bibfnamefont {M.~O.}\ \bibnamefont
  {Katanaev}},\ }\href@noop {} {\bibfield  {journal} {\bibinfo  {journal}
  {arXiv.org}\ } (\bibinfo {year} {2005})},\ \Eprint
  {http://arxiv.org/abs/cond-mat/0502123v1} {cond-mat/0502123v1} \BibitemShut
  {NoStop}%
\bibitem [{\citenamefont {Witten}(2016)}]{Witten2016}%
  \BibitemOpen
  \bibfield  {author} {\bibinfo {author} {\bibfnamefont {E.}~\bibnamefont
  {Witten}},\ }\href@noop {} {\bibfield  {journal} {\bibinfo  {journal}
  {Reviews of Modern Physics}\ }\textbf {\bibinfo {volume} {88}},\ \bibinfo
  {pages} {035001} (\bibinfo {year} {2016})}\BibitemShut {NoStop}%
\bibitem [{\citenamefont {Kellermann}\ \emph {et~al.}(2019)\citenamefont
  {Kellermann}, \citenamefont {Schmidt},\ and\ \citenamefont
  {Zimmer}}]{Kellermann19}%
  \BibitemOpen
  \bibfield  {author} {\bibinfo {author} {\bibfnamefont {N.}~\bibnamefont
  {Kellermann}}, \bibinfo {author} {\bibfnamefont {M.}~\bibnamefont {Schmidt}},
  \ and\ \bibinfo {author} {\bibfnamefont {F.~M.}\ \bibnamefont {Zimmer}},\
  }\href {\doibase 10.1103/PhysRevE.99.012134} {\bibfield  {journal} {\bibinfo
  {journal} {Phys. Rev. E}\ }\textbf {\bibinfo {volume} {99}},\ \bibinfo
  {pages} {012134} (\bibinfo {year} {2019})}\BibitemShut {NoStop}%
\bibitem [{\citenamefont {Schmidt}\ \emph {et~al.}(2020)\citenamefont
  {Schmidt}, \citenamefont {Kellermna},\ and\ \citenamefont
  {Zimmer}}]{Schmidt20}%
  \BibitemOpen
  \bibfield  {author} {\bibinfo {author} {\bibfnamefont {M.}~\bibnamefont
  {Schmidt}}, \bibinfo {author} {\bibfnamefont {N.}~\bibnamefont {Kellermna}},
  \ and\ \bibinfo {author} {\bibfnamefont {F.~M.}\ \bibnamefont {Zimmer}},\
  }\href@noop {} {\bibfield  {journal} {\bibinfo  {journal} {Phys. Rev. E}\
  }\textbf {\bibinfo {volume} {102}},\ \bibinfo {pages} {032138} (\bibinfo
  {year} {2020})}\BibitemShut {NoStop}%
\bibitem [{\citenamefont {Narayan}\ \emph {et~al.}(2019)\citenamefont
  {Narayan}, \citenamefont {Cano}, \citenamefont {Balatsky},\ and\
  \citenamefont {Spaldin}}]{Narayan19}%
  \BibitemOpen
  \bibfield  {author} {\bibinfo {author} {\bibfnamefont {A.}~\bibnamefont
  {Narayan}}, \bibinfo {author} {\bibfnamefont {A.}~\bibnamefont {Cano}},
  \bibinfo {author} {\bibfnamefont {A.}~\bibnamefont {Balatsky}}, \ and\
  \bibinfo {author} {\bibfnamefont {N.}~\bibnamefont {Spaldin}},\ }\href@noop
  {} {\bibfield  {journal} {\bibinfo  {journal} {Nature Materials}\ }\textbf
  {\bibinfo {volume} {18}},\ \bibinfo {pages} {223} (\bibinfo {year}
  {2019})}\BibitemShut {NoStop}%
\bibitem [{\citenamefont {Chandra}(2019)}]{Chandra19}%
  \BibitemOpen
  \bibfield  {author} {\bibinfo {author} {\bibfnamefont {P.}~\bibnamefont
  {Chandra}},\ }\href@noop {} {\bibfield  {journal} {\bibinfo  {journal}
  {Nature Materials}\ }\textbf {\bibinfo {volume} {18}},\ \bibinfo {pages}
  {197} (\bibinfo {year} {2019})}\BibitemShut {NoStop}%
\bibitem [{\citenamefont {Handunkanda}\ \emph {et~al.}(2015)\citenamefont
  {Handunkanda}, \citenamefont {Curry}, \citenamefont {Voronov}, \citenamefont
  {Said}, \citenamefont {Guzm\'an-Verri}, \citenamefont {Brierley},
  \citenamefont {Littlewood},\ and\ \citenamefont {Hancock}}]{Hancock15}%
  \BibitemOpen
  \bibfield  {author} {\bibinfo {author} {\bibfnamefont {S.~U.}\ \bibnamefont
  {Handunkanda}}, \bibinfo {author} {\bibfnamefont {E.~B.}\ \bibnamefont
  {Curry}}, \bibinfo {author} {\bibfnamefont {V.}~\bibnamefont {Voronov}},
  \bibinfo {author} {\bibfnamefont {A.~H.}\ \bibnamefont {Said}}, \bibinfo
  {author} {\bibfnamefont {G.~G.}\ \bibnamefont {Guzm\'an-Verri}}, \bibinfo
  {author} {\bibfnamefont {R.~T.}\ \bibnamefont {Brierley}}, \bibinfo {author}
  {\bibfnamefont {P.~B.}\ \bibnamefont {Littlewood}}, \ and\ \bibinfo {author}
  {\bibfnamefont {J.~N.}\ \bibnamefont {Hancock}},\ }\href {\doibase
  10.1103/PhysRevB.92.134101} {\bibfield  {journal} {\bibinfo  {journal} {Phys.
  Rev. B}\ }\textbf {\bibinfo {volume} {92}},\ \bibinfo {pages} {134101}
  (\bibinfo {year} {2015})}\BibitemShut {NoStop}%
\bibitem [{\citenamefont {Schmiedeshoff}\ \emph {et~al.}(2011)\citenamefont
  {Schmiedeshoff}, \citenamefont {Mun}, \citenamefont {Lounsbury},
  \citenamefont {Tracy}, \citenamefont {Palm}, \citenamefont {Hannahs},
  \citenamefont {Park}, \citenamefont {Murphy}, \citenamefont {Budko},\ and\
  \citenamefont {Canfield}}]{Schmiedeshoff11}%
  \BibitemOpen
  \bibfield  {author} {\bibinfo {author} {\bibfnamefont {G.}~\bibnamefont
  {Schmiedeshoff}}, \bibinfo {author} {\bibfnamefont {E.}~\bibnamefont {Mun}},
  \bibinfo {author} {\bibfnamefont {A.}~\bibnamefont {Lounsbury}}, \bibinfo
  {author} {\bibfnamefont {S.}~\bibnamefont {Tracy}}, \bibinfo {author}
  {\bibfnamefont {E.}~\bibnamefont {Palm}}, \bibinfo {author} {\bibfnamefont
  {S.}~\bibnamefont {Hannahs}}, \bibinfo {author} {\bibfnamefont {J.-H.}\
  \bibnamefont {Park}}, \bibinfo {author} {\bibfnamefont {T.}~\bibnamefont
  {Murphy}}, \bibinfo {author} {\bibfnamefont {S.}~\bibnamefont {Budko}}, \
  and\ \bibinfo {author} {\bibfnamefont {P.}~\bibnamefont {Canfield}},\ }\href
  {https://link.aps.org/doi/10.1103/PhysRevB.83.180408} {\bibfield  {journal}
  {\bibinfo  {journal} {Phys. Rev. B}\ }\textbf {\bibinfo {volume} {83}},\
  \bibinfo {pages} {180408} (\bibinfo {year} {2011})}\BibitemShut {NoStop}%
\bibitem [{\citenamefont {Tokiwa}\ \emph {et~al.}(2013)\citenamefont {Tokiwa},
  \citenamefont {Garst}, \citenamefont {Gegenwart}, \citenamefont {Bud/ko},\
  and\ \citenamefont {Canfield}}]{Tokawa13}%
  \BibitemOpen
  \bibfield  {author} {\bibinfo {author} {\bibfnamefont {Y.}~\bibnamefont
  {Tokiwa}}, \bibinfo {author} {\bibfnamefont {M.}~\bibnamefont {Garst}},
  \bibinfo {author} {\bibfnamefont {P.}~\bibnamefont {Gegenwart}}, \bibinfo
  {author} {\bibfnamefont {S.}~\bibnamefont {Bud/ko}}, \ and\ \bibinfo {author}
  {\bibfnamefont {P.}~\bibnamefont {Canfield}},\ }\href
  {https://link.aps.org/doi/10.1103/PhysRevLett.111.116401} {\bibfield
  {journal} {\bibinfo  {journal} {Phys. Rev. Lett.}\ }\textbf {\bibinfo
  {volume} {111}},\ \bibinfo {pages} {116401} (\bibinfo {year}
  {2013})}\BibitemShut {NoStop}%
\bibitem [{\citenamefont {Steppke}\ \emph {et~al.}(2013)\citenamefont
  {Steppke}, \citenamefont {Kuchler}, \citenamefont {Lausberg}, \citenamefont
  {Lengyel}, \citenamefont {Steinke}, \citenamefont {Borth}, \citenamefont
  {Luhmann}, \citenamefont {Krellner}, \citenamefont {Nicklas}, \citenamefont
  {Geibel}, \citenamefont {Steglich},\ and\ \citenamefont
  {Brando}}]{Steppke13}%
  \BibitemOpen
  \bibfield  {author} {\bibinfo {author} {\bibfnamefont {A.}~\bibnamefont
  {Steppke}}, \bibinfo {author} {\bibfnamefont {R.}~\bibnamefont {Kuchler}},
  \bibinfo {author} {\bibfnamefont {S.}~\bibnamefont {Lausberg}}, \bibinfo
  {author} {\bibfnamefont {E.}~\bibnamefont {Lengyel}}, \bibinfo {author}
  {\bibfnamefont {L.}~\bibnamefont {Steinke}}, \bibinfo {author} {\bibfnamefont
  {R.}~\bibnamefont {Borth}}, \bibinfo {author} {\bibfnamefont
  {T.}~\bibnamefont {Luhmann}}, \bibinfo {author} {\bibfnamefont
  {C.}~\bibnamefont {Krellner}}, \bibinfo {author} {\bibfnamefont
  {M.}~\bibnamefont {Nicklas}}, \bibinfo {author} {\bibfnamefont
  {C.}~\bibnamefont {Geibel}}, \bibinfo {author} {\bibfnamefont
  {F.}~\bibnamefont {Steglich}}, \ and\ \bibinfo {author} {\bibfnamefont
  {M.}~\bibnamefont {Brando}},\ }\href
  {http://science.sciencemag.org/content/339/6122/933} {\bibfield  {journal}
  {\bibinfo  {journal} {Science}\ }\textbf {\bibinfo {volume} {339}},\ \bibinfo
  {pages} {933} (\bibinfo {year} {2013})}\BibitemShut {NoStop}%
\bibitem [{\citenamefont {Pikin}(1970)}]{Pikin70}%
  \BibitemOpen
  \bibfield  {author} {\bibinfo {author} {\bibfnamefont {S.~A.}\ \bibnamefont
  {Pikin}},\ }\href@noop {} {\bibfield  {journal} {\bibinfo  {journal} {Sov.
  Phys. JETP}\ }\textbf {\bibinfo {volume} {31}},\ \bibinfo {pages} {753}
  (\bibinfo {year} {1970})}\BibitemShut {NoStop}%
\bibitem [{\citenamefont {Quader}\ and\ \citenamefont
  {Widom}(2014)}]{Quader14}%
  \BibitemOpen
  \bibfield  {author} {\bibinfo {author} {\bibfnamefont {K.}~\bibnamefont
  {Quader}}\ and\ \bibinfo {author} {\bibfnamefont {M.}~\bibnamefont {Widom}},\
  }\href {https://link.aps.org/doi/10.1103/PhysRevB.90.144512} {\bibfield
  {journal} {\bibinfo  {journal} {Phys. Rev. B}\ }\textbf {\bibinfo {volume}
  {90}},\ \bibinfo {pages} {144512} (\bibinfo {year} {2014})}\BibitemShut
  {NoStop}%
\bibitem [{\citenamefont {Chandra}\ \emph {et~al.}(2013)\citenamefont
  {Chandra}, \citenamefont {Coleman},\ and\ \citenamefont {Flint}}]{Chandra13}%
  \BibitemOpen
  \bibfield  {author} {\bibinfo {author} {\bibfnamefont {P.}~\bibnamefont
  {Chandra}}, \bibinfo {author} {\bibfnamefont {P.}~\bibnamefont {Coleman}}, \
  and\ \bibinfo {author} {\bibfnamefont {R.}~\bibnamefont {Flint}},\ }\href
  {https://www.nature.com/articles/nature11820} {\bibfield  {journal} {\bibinfo
   {journal} {Nature}\ }\textbf {\bibinfo {volume} {493}},\ \bibinfo {pages}
  {621} (\bibinfo {year} {2013})}\BibitemShut {NoStop}%
\end{thebibliography}%


\end{document}